\documentclass[twocolumn,amsmath,amssymb,nofootinbib,superscriptaddress,longbibliography,floatfix]{revtex4-1}

\usepackage{graphicx}
\usepackage{amsmath}
\usepackage{amssymb}
\usepackage{xcolor}
\usepackage{hyperref}
\hypersetup{pdfstartview={FitH},pdfpagemode={UseNone}, breaklinks=true, colorlinks=true, linkcolor=blue, citecolor=blue, urlcolor=blue, bookmarksopen=true, pdfnewwindow=true}
\usepackage[all]{hypcap}
\usepackage[english]{babel}
\usepackage{physics}
\usepackage{braket}

\begin{document}

\preprint{APS/123-QED}

\title{Theoretical description of interface states in a tetragonal lattice of bianisotropic resonators}

\author{Alina D. Rozenblit}
\email{alina.rozenblit@metalab.ifmo.ru}
\affiliation{School of Physics and Engineering, ITMO University, 197101 Saint Petersburg, Russia}

\author{Nikita A. Olekhno}
\affiliation{School of Physics and Engineering, ITMO University, 197101 Saint Petersburg, Russia}

\date{\today}

\begin{abstract} 
In the present paper, we construct a theoretical description of a three-dimensional photonic structure in the form of a tetragonal lattice of bianisotropic resonators applying a dyadic Green's function approach. By representing the resonators as point electric and magnetic dipoles, we obtain the Bloch Hamiltonians for the approximations considering the interactions between the nearest, next-nearest, and next-to-next-nearest resonators, and construct the corresponding real-space tight-binding models. We analyze the band diagrams, spatial structure of the eigenmodes, and their localization, revealing quadratic degeneracies in the vicinity of high-symmetry points in the absence of bianisotropy and the emergence of in-gap states localized at a domain wall upon the introduction of bianisotropy. Finally, we compare the theoretical results with full-wave numerical simulations for an array of bianisotropic resonators.
\end{abstract}

\maketitle

\section{Introduction}
\label{sec:Introduction}

Photonic topological insulators (PTIs)~\cite{2014_Lu,2019_Ozawa} provide a universal framework for the localization of electromagnetic fields at frequencies within the photonic band gap~\cite{2026_Leykam}. In addition to frequency separation from the bulk excitations, such states gain additional robustness against various imperfections in the PTI structure due to symmetry protection~\cite{2019_Kruk}. While one-dimensional (1D) structures~\cite{2019_Kruk,2021_Bobylev} and two-dimensional (2D) metasurfaces~\cite{2018_Gorlach,2025_Rozenblit} appear more compact and thus more suitable for fabrication and applications~\cite{2020_Ma,2023_Puchnin}, three-dimensional (3D) PTIs demonstrate a rich hierarchy of surface, hinge, and corner states~\cite{2025_Wang}, providing great opportunities to study novel physics such as axion insulators~\cite{2025_Liu,2025_Lai} and implement light guiding over complex trajectories~\cite{2025_Lai}.

There are various strategies for opening a topological band gap that range from dimerized~\cite{1979_Su} or shrink-expanded~\cite{2015_Wu} models that preserve time reversal symmetry to the incorporation of magnetic field in the model, either directly via magneto-optical effects~\cite{2009_Wang} or effectively by time modulation~\cite{2012_Fang}. Depending on the considered mechanisms of symmetry breaking that remove spectral degeneracies and result in the opening of a band gap, such PTIs can represent, for example, analogs of Chern insulators (demonstrating the quantum Hall effect) with chiral edge states and broken time-reversal symmetry~\cite{2008_Haldane}, time-reversal analogs to quantum spin Hall~\cite{2013_Khanikaev, 2020_Xie} or valley Hall~\cite{2018_Noh, 2018_Gao} insulators with the propagation direction of edge states defined by pseudospin or valley degrees of freedom, or higher-order topological insulators~\cite{2017_Benalcazar_Science, 2019_Xie, 2019_Chen, 2020_Li, 2025_Wang}.

Recently, the introduction of a bianisotropic response at the level of a single resonator was shown to result in a topological band gap in 1D chains of such resonators~\cite{2019_Gorlach}, as well as 2D~\cite{2019_Slobozhanyuk, 2025_Rozenblit} and 3D~\cite{2017_Slobozhanyuk, 2026_Zhirihin} arrays of evenly spaced resonators arranged in simple lattices. Physically, the bianisotropic response achieved via the breaking of the spatial inversion symmetry of the resonators results in the mixing of electric and magnetic modes that is analogous to the spin-orbit interaction~\cite{2013_Khanikaev}. Hexagonal-lattice based 2D~\cite{2016_Cheng, 2019_Slobozhanyuk} and 3D~\cite{2017_Slobozhanyuk, 2019_Yang, 2026_Zhirihin} resonator arrays that demonstrate linear degeneracies in high-symmetry points of the Brillouin zone in the absence of bianisotropy and can be described by effective Dirac Hamiltonians are well-studied theoretically and have been implemented experimentally. However, their square-lattice based counterparts with $C_{4}$ rotational symmetry and quadratic degeneracies~\cite{2008_Chong} remain less explored. In particular, 2D square lattice arrays have recently been implemented~\cite{2016_Slobozhanyuk, 2025_Rozenblit}, while a 3D simple cubic lattice has been analyzed within the perturbation theory framework~\cite{2017_Ochiai}, which allows describing the band structure only in the vicinity of high-symmetry points. In addition, bianisotropic resonator arrays can be homogenized and represented as an effective bianisotropic medium~\cite{2019_Guo, 2022_Chern}, although such a description may not fully account for the symmetry of the lattice.

\begin{figure*}[t]
  \centering
  \includegraphics[width=16cm]{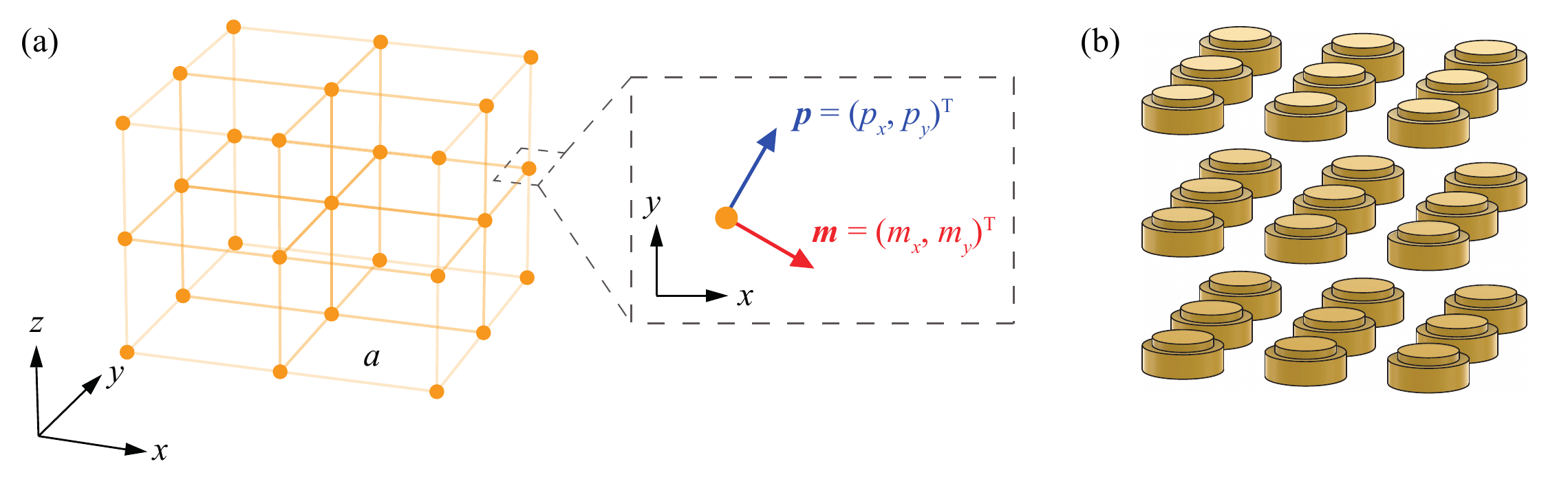}
  \caption{(a) The schematic of point electric and magnetic dipoles arranged in a cubic lattice with period $a$. The inset shows the orientation plane of the dipoles $\mathbf{p}$ and $\mathbf{m}$. (b) The example of the model realization as an array of dielectric resonators with inversion symmetry broken shape.}
  \label{fig:System}
\end{figure*}

In the present paper, we apply the dyadic Green's function approach~\cite{2019_Gorlach, 2025_Rozenblit} to further advance these results and develop a theoretical model Fig.~\ref{fig:System}(a) of a tetragonal lattice of bianisotropic resonators shown in Fig.~\ref{fig:System}(b) that allows one to calculate the band structure in the entire Brillouin zone and visualize the eigenmodes in finite arrays. To study the effects of long-range couplings, we consider three different approximations that include the couplings between the nearest, the next-nearest, and the next-to-next nearest resonators. With the help of the obtained model, we analyze the effects of the bianisotropic response magnitude on the band structures of the considered resonator arrays.

The paper is organized as follows. In Section~\ref{sec:Theory}, we introduce the approximations, derive Bloch Hamiltonians, and construct real-space tight-binding models for a cubic lattice of bianisotropic resonators using dyadic Green's functions. In Section~\ref{sec:Dispersion}, we analyze band diagrams and the density of states. In Section~\ref{sec:Localization}, we study the localization properties and spatial profiles of eigenstates in the finite model, and visualize the band structure of the interface states. Section~\ref{sec:Simulations} considers numerical simulations of an array of bianisotropic resonators and its comparison with theoretical results. Finally, Section~\ref{sec:Discussion} contains a discussion of the results.

\section{Theoretical framework}
\label{sec:Theory}

In this Section, we extend the coupled-dipole method previously applied to describe a one-dimensional chain~\cite{2019_Gorlach} and a square lattice~\cite{2025_Rozenblit} of dielectric resonators with a bianisotropic response to the three-dimensional case.

\subsection{Considered approximations}
\label{sec:Approximations}

In the construction of theoretical models, we will rely on the following approximations. We assume that each of the particles in the array supports hybridization of the electric and magnetic dipole moments oriented in the $xy$-plane, while the electromagnetic $z$-component can be neglected~\cite{2025_Rozenblit}. Thus, we consider point electric ${\mathbf{p} = (p_{x}, p_{y})^{\rm T}}$ and magnetic $\mathbf{m} = (m_{x}, m_{y})^{\rm T}$ dipoles oriented in the $xy$-plane and arranged in a cubic lattice with period $a$, as shown in Fig.~\ref{fig:System}(a). The components of the excited electric $\mathbf{E} = (E_{x}, E_{y})^{\rm T}$ and magnetic $\mathbf{H} = (H_{x}, H_{y})^{\rm T}$ fields and induced dipole moments in a lattice node with coordinates $(ia, ja, ka)$ are related by the polarizability tensor $\widehat{\alpha}$:
\begin{equation}
    \begin{pmatrix}
    \mathbf{p}^{ijk}\\
    \mathbf{m}^{ijk}
    \end{pmatrix}
    = 
    \widehat{\alpha}
    \begin{pmatrix}
    \mathbf{E}^{ijk}\\
    \mathbf{H}^{ijk}
    \end{pmatrix}
    =
    \begin{pmatrix}
    \widehat{\alpha}^{\rm ee} & \widehat{\alpha}^{\rm em}\\
    \widehat{\alpha}^{\rm me} & \widehat{\alpha}^{\rm mm}
    \end{pmatrix}
    \begin{pmatrix}
    \mathbf{E}^{ijk}\\
    \mathbf{H}^{ijk}
    \end{pmatrix}.
    \label{eq:Dipole_moments}
\end{equation}
which includes the electric $\widehat{\alpha}^{\rm ee}$, magnetic $\widehat{\alpha}^{\rm mm}$, electromagnetic $\widehat{\alpha}^{\rm em}$, and magnetoelectric $\widehat{\alpha}^{\rm me}$ components.

First, along with the point dipole approximation, we apply a near-field (quasi-static) approximation when describing the resonators that form an array, which is valid only when the characteristic size of individual resonators $d$, as well as the spacing between the nearest resonators $a$, are much lower that the light wavelength at the frequencies of interest: $k_{0}a,k_{0}d \ll 1$, with $k_{o}$ being the wave number of light in vacuum. Within such an approximation, only the near-field terms proportional to $1/r^3$ remain, where $r$ is the distance between the considered lattice nodes. Moreover, since we assume that only electric and magnetic dipoles in the $xy$ plane in Fig.~\ref{fig:System} are excited in the considered frequency region, the height of resonators should be different from their diameter, so that the frequencies of dipole modes along the $z$-axis are separated from the frequencies of dipole modes in the $xy$ plane. The considered approximations are satisfied for the ceramic resonators used in experiments with 2D~\cite{2019_Slobozhanyuk, 2025_Rozenblit} and 3D~\cite{2026_Zhirihin} structures.

Second, we consider the case when the electromagnetic duality ($\widehat{\varepsilon} = \widehat{\mu}$) is satisfied, resulting in equal electric and magnetic polarizabilities $\beta$ along the $x$- and $y$-axes $\widehat{\alpha}^{\rm ee} = \widehat{\alpha}^{\rm mm} = \beta \cdot\widehat{\sigma}_0$, where $\widehat{\sigma}_0 = (1,0;0,1)$ is the unity matrix. The electromagnetic duality leads to the degeneracy of electric and magnetic dipole modes at the high-symmetry points that is analogous to Kramers pair degeneracy in quantum spin Hall insulators~\cite{2013_Khanikaev}. This effect can be achieved by tuning the shape of the resonator~\cite{2016_Cheng, 2016_Slobozhanyuk, 2019_Slobozhanyuk, 2019_Yang}. At the same time, we consider reciprocal bianisotropic particles demonstrating the presence of non-vanishing electromagnetic (magnetoelectric) components described by the electromagnetic coupling $\chi$~\cite{2018_Asadchy} $\widehat{\alpha}^{\rm em} = -\{\widehat{\alpha}^{\rm me}\}^{\rm T} = -\chi \cdot \widehat{\sigma}_2$, where $\widehat{\sigma}_2 = (0, -{\rm i}; {\rm i}, 0)$ is the Pauli matrix. We also assume that the diagonal elements of the inverse polarizability tensor
\begin{equation}
    \widehat{\alpha}^{-1}
    = \dfrac{1}{\beta^2-\chi^2}
    \begin{pmatrix}
        \beta & 0 & 0 & -{\rm i}\chi\\
        0 & \beta & {\rm i}\chi & 0\\
        0 & -{\rm i}\chi & \beta & 0\\
        {\rm i}\chi & 0 & 0 & \beta
    \end{pmatrix}
    \label{Inverse_polarizability_tensor_main}
\end{equation}
depend on the resonance frequency $f_0$ as ${\beta/(\beta^2-\chi^2) \propto f - f_0}$, while the term $\chi/(\beta^2-\chi^2)$ can be approximated as a constant~\cite{2023_Kim}.

\subsection{Effective Bloch Hamiltonian}
\label{sec:Derivation}

To construct the theoretical model, we consider the electric and magnetic field components at the lattice site with coordinates $(ia,ja,ka)$, where $a$ is the lattice constant, that can be expressed through dipole moments with the help of the inverse polarizability tensor $\widehat{\alpha}^{-1}$:
\begin{equation}
    \begin{pmatrix}
    \mathbf{E}^{ijk}\\
    \mathbf{H}^{ijk}
    \end{pmatrix} =
    \widehat{\alpha}^{-1}
    \begin{pmatrix}
    \mathbf{p}^{ijk}\\
    \mathbf{m}^{ijk}
    \end{pmatrix}.
    \label{eq:Fields_by_dipole_moments_main}
\end{equation}
On the other hand, the same fields can be expressed as the sum of the fields excited by dipoles from other sites:
\begin{equation}
    \begin{pmatrix}
        \mathbf{E}^{ijk} \\
        \mathbf{H}^{ijk}
    \end{pmatrix} = \sum_{\substack{m \ne i\\
    n \ne j \\ l \ne k}}
    \begin{pmatrix}
        \widehat{G}^{\rm ee}(r,k_0) & \widehat{G}^{\rm em}(r,k_0)\\
        \widehat{G}^{\rm me}(r,k_0) & \widehat{G}^{\rm mm}(r,k_0)
    \end{pmatrix}
        \begin{pmatrix}
        \mathbf{p}^{mnl} \\
        \mathbf{m}^{mnl}
    \end{pmatrix},
    \label{eq:E_and_Green_main}
\end{equation}
where $r = a\sqrt{(i-m)^2+(j-n)^2+(k-l)^2}$ is the distance between the lattice sites with coordinates $(ma,na,la)$ and $(ia,ja,ka)$, and $k_0$ is the wave number. The bianisotropic parts of the dyadic Green's function are related as $\widehat{G}^{\rm em}(r, k_0)=-\widehat{G}^{\rm me}(r, k_0)$, while the electromagnetic duality requires $\widehat{G}^{\rm ee}(r, k_0)=\widehat{G}^{\rm mm}(r, k_0)$. Dyadic Green's functions for point dipoles can be calculated using the following equations~\cite{1994_Tai}:
\begin{align}
    G^{\rm ee}_{\zeta\eta} & = (\partial_{\zeta} \partial_{\eta} + k^2_0\delta_{\zeta\eta})\dfrac{e^{{\rm i} k_0 r_{ij}}}{r_{ij}},
    \label{eq:Gee} \\
    G^{\rm em}_{\zeta\eta} & = {\rm i} k_0 \varepsilon_{\zeta\eta\kappa}\partial_{\kappa}\dfrac{e^{{\rm i} k_0 r_{ij}}}{r_{ij}},
    \label{eq:Gem}
\end{align}
where $\delta_{\zeta\eta}$ is the Kronecker delta, $\varepsilon_{\zeta \eta \kappa}$ is the Levi-Civita symbol, and $\partial_\zeta = \partial/\partial\zeta$. The parameters $\zeta$, $\eta$, $\kappa$ take the values $\{x,y,z\}$.

In order to study the effects of long-range interactions that may sufficiently contribute to the properties of photonic structures~\cite{2020_Li, 2022_D4}, we consider three cases: the model with interactions only between the nearest sites separated by the distance $r=a$ (Model~I) the model which includes the couplings between the nearest ($r=a$) and the next-nearest ($r = \sqrt{2}a$) sites (Model~II), as well as the model which takes into account the interactions between the sites in the first ($r=a$), the second ($r = \sqrt{2}a$), and the third ($r=\sqrt{3}a$) coordination spheres (Model~III). Then, applying Bloch's theorem to Eq.~\eqref{eq:E_and_Green_main} and comparing the result with Eq.~\eqref{eq:Fields_by_dipole_moments_main}, we obtain the eigenvalue problem $\widehat{H}\ket{\psi}=\lambda\ket{\psi}$ in the pseudospin basis $\ket{\psi}=(p_x + m_x, p_y + m_y, p_x - m_x, p_y - m_y)^{\rm T}$~\cite{2017_Slobozhanyuk} with eigenvalues {$\lambda = a^3\beta / (\beta^2-\chi^2)$} and the bianisotropy parameter {$\Omega = a^3 \chi / (\beta^2-\chi^2)$}. The full Bloch Hamiltonian $\widehat{H}$ has the block-diagonal form
\begin{equation}
    \widehat{H} = 
    \begin{pmatrix}
        \widehat{H}^{\uparrow} & 0\\
        0 & \widehat{H}^{\downarrow}
    \end{pmatrix},
 \label{eq:H_I_spin_basis_full}
\end{equation}
where $\widehat{H}^{\uparrow(\downarrow)}$ are Bloch Hamiltonians for two pseudospins denoted as $\uparrow(\downarrow)$, respectively.
For the three considered approximations in dimensionless units $a=1$,
\begin{multline}
    \widehat{H}^{\uparrow(\downarrow)}_{\rm I} =
    (\cos{k_x}+\cos{k_y}-2\cos{k_z}) \widehat{\sigma}_0 +\\
    +3(\cos{k_x}-\cos{k_y}) \widehat{\sigma}_3 \pm \Omega \widehat{\sigma}_2
    \label{eq:H_I_spin_basis_component}
\end{multline}
for Model~I,
\begin{multline}
    \widehat{H}_{\rm II}^{\uparrow(\downarrow)} = \Big(\cos{k_x}+\cos{k_y}-2\cos{k_z} +\\
    + \dfrac{2\cos{k_x} \cos{k_y} - \cos{k_x} \cos{k_z} - \cos{k_y} \cos{k_z}}{2\sqrt{2}}\Big) \widehat{\sigma}_0  + \\
    + 3\Big(\cos{k_x}-\cos{k_y} + \dfrac{\cos{k_x} \cos{k_z}-\cos{k_y} \cos{k_z}}{2\sqrt{2}}\Big)  \widehat{\sigma}_3 -\\
    - \dfrac{3}{\sqrt{2}} \sin{k_x} \sin{k_y} \widehat{\sigma}_1 \pm \Omega \widehat{\sigma}_2
    \label{eq:H_II_spin_basis_component}
\end{multline}
for Model~II, and
\begin{multline}
    \widehat{H}_{\rm III}^{\uparrow(\downarrow)}
    = \Big(\cos{k_x}+\cos{k_y}-2\cos{k_z} +\\
    + \dfrac{2\cos{k_x} \cos{k_y} - \cos{k_x} \cos{k_z} - \cos{k_y} \cos{k_z}}{2\sqrt{2}}\Big) \widehat{\sigma}_0 + \\
    + 3\Big(\cos{k_x}-\cos{k_y} + \dfrac{\cos{k_x} \cos{k_z}-\cos{k_y} \cos{k_z}}{2\sqrt{2}}\Big) \widehat{\sigma}_3 -\\
    - \Big(\dfrac{3}{\sqrt{2}}+\dfrac{8}{3\sqrt{3}}\cos{k_z}\Big) \sin{k_x} \sin{k_y} \widehat{\sigma}_1 \pm \Omega \widehat{\sigma}_2
    \label{eq:H_III_spin_basis_component}
\end{multline}
for Model~III, where $\widehat{\sigma}_0 = (1,0;0,1)$ is the unity matrix and $\widehat{\sigma}_1 = (0,1;1,0)$, $\widehat{\sigma}_2 = (0, -{\rm i}; {\rm i}, 0)$, and $\widehat{\sigma}_3 = (1,0;0,-1)$ are the Pauli matrices.

The eigenvalues of pseudospin Hamiltonians are doubly-degenerate for the two values of pseudospin and have the following form for Eq.~\eqref{eq:H_I_spin_basis_component}:
\begin{multline}
    \lambda^{\uparrow(\downarrow)}_{\rm I} = \cos{k_x}+\cos{k_y}-2\cos{k_z} \pm \Big( \Omega^2 +  \frac{9}{2}(2 + \\ +\cos{(2k_x)} 
    + \cos{(2k_y)} - 4\cos{k_x}\cos{k_y})\Big)^{1/2}.
    \label{eq:Eigenvalues_I}
\end{multline}
The detailed derivation of Bloch Hamiltonians, as well as analytical expressions for the energy bands of pseudospin Hamiltonians Eq.~\eqref{eq:H_II_spin_basis_component} and Eq.~\eqref{eq:H_III_spin_basis_component}, are included in Supplemental Material~\cite{Supplement}.

\begin{figure*}[tbp]
    \centering
    \includegraphics[width=16cm]{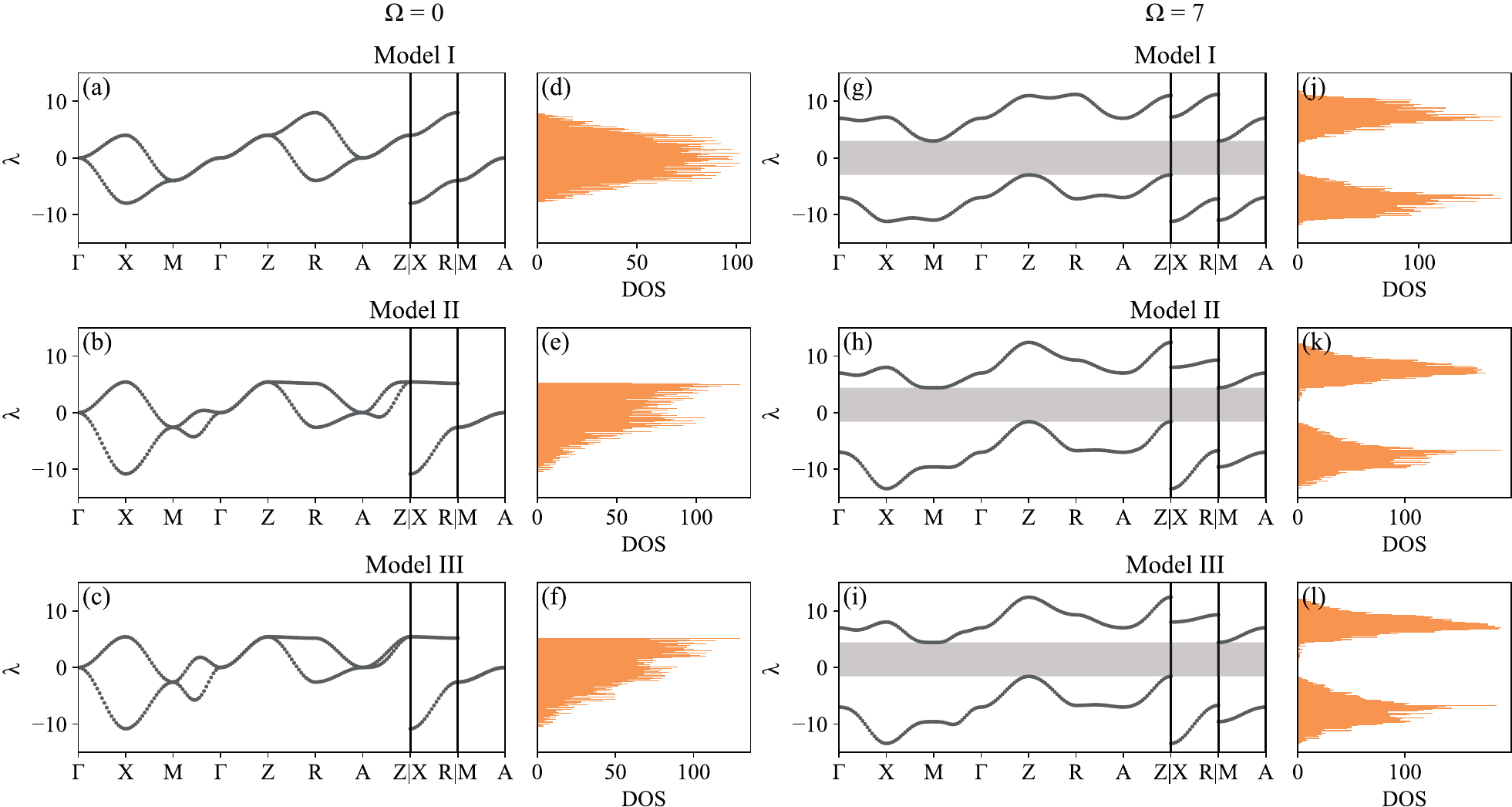}
    \caption{(a)-(c) Energy bands of the Bloch Hamiltonians for (a) Model I (the nearest neighbors), Eq.~\eqref{eq:H_I_spin_basis_component}; (b) Model II (the next-nearest neighbors), Eq.~\eqref{eq:H_II_spin_basis_component}; and (c) Model III (the next-to-next nearest neighbors), Eq.~\eqref{eq:H_III_spin_basis_component} in the absence of bianisotropy ($\Omega = 0$). (d)-(f) Density of states $\rho(\lambda)$ for tight-binding models corresponding to cubic lattices of $10 \times 20 \times 10$ sites without bianisotropy ($\Omega = 0$) for (d) Model I, (e) Model II, and (f) Model III real-space Hamiltonians plotted as histogram with 150 bins. Panels (g)-(l) are the same as (a)-(f), but for non-zero bianisotropy parameter $\Omega = 7$. The gray shaded areas in dispersion diagrams (g)-(i) highlight the bandgaps.}
    \label{fig:Dispersion_and_DOS}
\end{figure*}

\subsection{Real-space tight-binding models}
\label{sec:TB}

To study the spatial structure of eigenmodes in a finite lattice, we consider the real-space tight-binding model described by the following matrix~\cite{2023_Kim} (in units of $a = 1$):
\begin{equation}
     \widehat{B} = \widehat{M} \otimes \widehat{G}(r) - \widehat{\Omega} \otimes \widehat{\sigma}_1 \otimes \widehat{\sigma}_2,
     \label{eq:TB}
\end{equation}
where $\widehat{M}$ is the $[N \times N]$ connectivity matrix describing the connections between lattice sites with $N = N_{x}N_{y}N_{z}$ being the total number of sites in a cubic lattice: the matrix element $M_{st}=1$ if the sites labeled by indices $s$ and $t$ with coordinates $(x_s,y_s,z_s)$ and $(x_t,y_t,z_t)$ are connected with a link, $M_{st}=0$ otherwise, and the diagonal elements $M_{tt}=0$, and $\widehat{G}(r)$ is the dyadic Green's function in the near-field approximation that depends on the distance $r=r_{st}=((x_s-x_t)^2 + (y_s-y_t)^2 + (z_s-z_t)^2)^{1/2}$ between the sites $s$ and $t$. Thus, the first term on the right hand side of Eq.~\eqref{eq:TB} defines the interactions between the electric (magnetic) dipoles in different sites. The second term describes the hybridization of the electric and magnetic dipole moments in each of the lattice sites that is governed by the $[N \times N]$ bianisotropy parameter matrix $\Omega_{st}=\Omega(x_s,y_s,z_s)\delta_{st}$, where $\Omega(x_s,y_s,z_s)$ is the value of bianisotropy parameter in a certain site $s$ with coordinates $(x_s,y_s,z_s)$. We consider the following coordinate dependence of the bianisotropy parameter in a lattice with $2N_{x} \times N_{y} \times N_{z}$ sites:
\begin{equation}
    \Omega(x,y,z) = \Omega (x) = \begin{cases}
        \Omega & \text{, $x < N_x$} \\
        -\Omega & \text{, $x \geq N_x$}
    \end{cases}.
\end{equation}
This configuration corresponds to a domain wall between two cubic sublattices formed by $N_{x} \times N_{y} \times N_{z}$ resonators with opposite signs of the bianisotropy parameter for the domains.

\section{Spectral properties}
\label{sec:Dispersion}

\begin{figure*}[tbp]
    \centering
    \includegraphics[width=16cm]{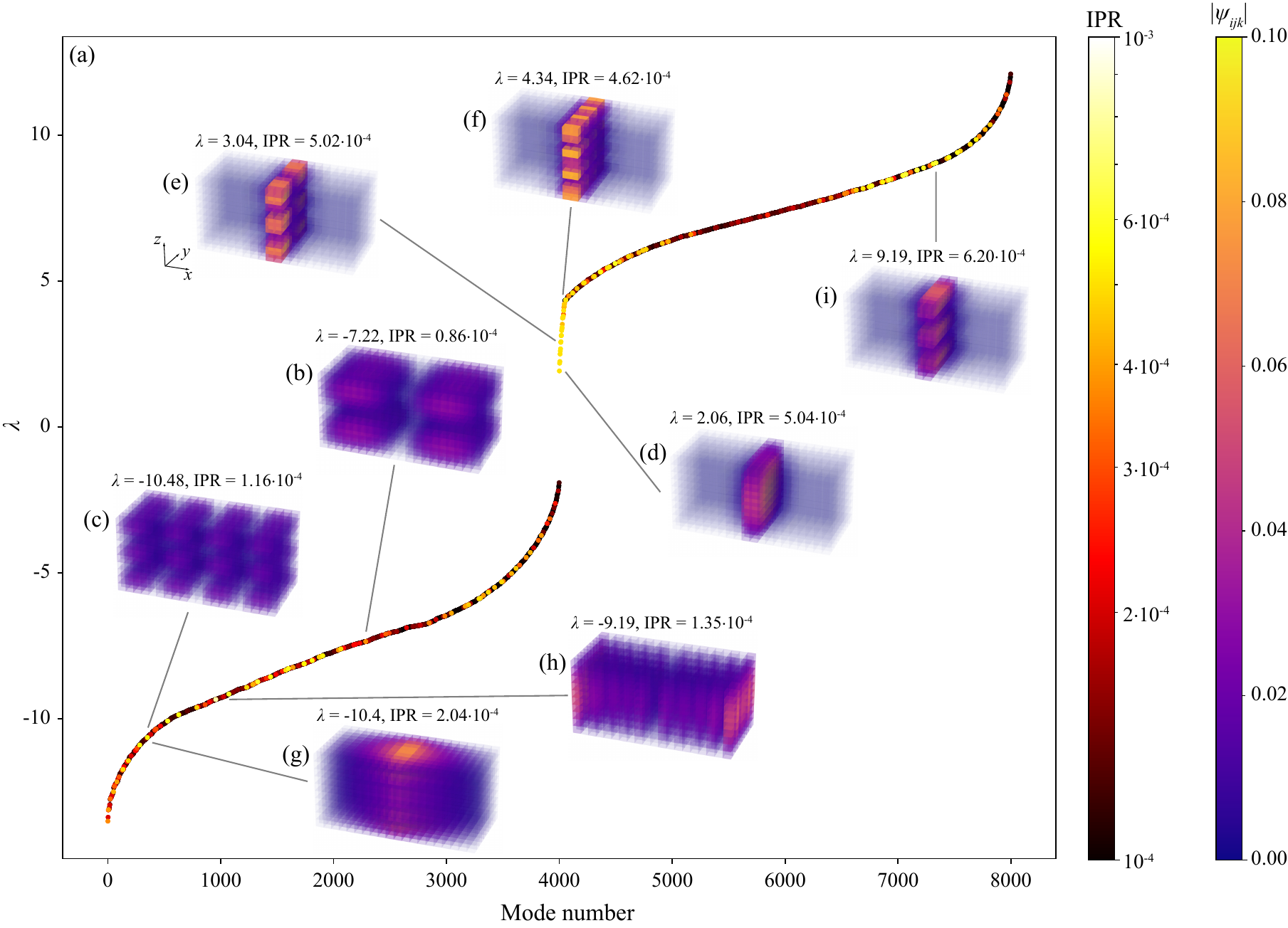}
    \caption{(a) Spectrum of eigenvalues $\lambda$ for the system consisting of two domains with $10 \times 10 \times 10$ sites and bianisotropy parameters $\Omega = 7$ and $\Omega = -7$, respectively. Color shows the inverse participation ratio Eq.~\eqref{eq:IPR}. (b)-(i) Eigenfunctions profiles corresponding to the absolute value of pseudospin-up polarization $p_x + m_x$ shown by color, which demonstrate (b),(c) bulk states, (d)-(f) interface states localized at the domain wall in the bandgap, (g) an interface state hybridized with a bulk mode, (h) a bulk mode with dominant surface localization at the boundary, and (i) an edge state in the continuum.}
    \label{fig:Localization}
\end{figure*}

The obtained pseudospin Bloch Hamiltonians $\widehat{H}^{\uparrow(\downarrow)}$ each have two eigenvalue bands, whereas the eigenvalues of the full Bloch Hamiltonians $\widehat{H}$ that correspond to the opposite pseudospins come in doubly-degenerate Kramers pairs~\cite{2007_Fu}. Figures~\ref{fig:Dispersion_and_DOS}(a)--~\ref{fig:Dispersion_and_DOS}(c) show the band diagrams in the Brillouin zone in the absence of bianisotropy ($\Omega = 0$) for each of the considered Models~I, II, and III, respectively. The eigenvalues are plotted along the $\Gamma(0,0,0)$-$X(\pi,0,0)$-$M(\pi,\pi,0)$-$\Gamma(0,0,0)$-$Z(0,0,\pi)$-$R(\pi,0,\pi)$-$A(\pi,\pi,\pi)$-$Z(0,0,\pi)$ trajectory, with additional insets showing the dispersion between $X$-$R$ and $M$-$A$ high-symmetry points.

It is seen that in the absence of bianisotropy, the bands for Model~I host fourfold-degenerate nodal lines between the points $M$-$\Gamma$, $\Gamma$-$Z$, $A$-$Z$, and $M$-$A$, Fig.~\ref{fig:Dispersion_and_DOS}(a). However, once the interactions between the next-nearest sites are taken into account, the degeneracies between high-symmetry points $M$-$\Gamma$ and $A$-$Z$ are lifted, as demonstrated in Fig.~\ref{fig:Dispersion_and_DOS}(b). Additional couplings between the lattice sites in the third coordination sphere only slightly change the dispersion between the diagonally-opposite high-symmetry points $M$-$\Gamma$ and $A$-$Z$, Fig.~\ref{fig:Dispersion_and_DOS}(c). Importantly, quadratic fourfold degeneracies are observed at high-symmetry points $\Gamma$, $M$, $Z$, and $A$ in all the considered models, in contrast to the linear Dirac-like degeneracies characteristic of 3D resonator arrays based on stacked hexagonal lattices~\cite{2017_Slobozhanyuk} or their 2D counterparts~\cite{2019_Slobozhanyuk}, in analogy to a 2D square lattice of bianisotropic resonators~\cite{2025_Rozenblit}. Finally, Models~II and III feature well-pronounced flat bands between the points $Z$-$R$ and $X$-$R$ for the upper energy band, Figs.~\ref{fig:Dispersion_and_DOS}(b) and~\ref{fig:Dispersion_and_DOS}(c).

Along with band diagrams for Bloch Hamiltonians, we also numerically evaluate the eigenvalues for finite-size tight-binding lattices from Eq.~\eqref{eq:TB}. Then, we consider the density of states (DOS)
\begin{equation}
   \rho(\lambda) = \sum_{j}\delta(\lambda-\lambda_{j}),
   \label{eq:DOS}
\end{equation}
where the index $j$ enumerates all eigenvalues $\lambda_j$ and $\delta(\lambda-\lambda_j)$ is the pseudo-delta function defined in the following way: $\delta(\lambda-\lambda_j)=0$ for $\lambda \neq \lambda_j$ and $\delta(\lambda-\lambda_j)=1$ for $\lambda = \lambda_j$. Numerically, the DOS can be obtained by plotting a histogram of the evaluated eigenvalues. In Figs.~\ref{fig:Dispersion_and_DOS}(d)-~\ref{fig:Dispersion_and_DOS}(f), the density of states $\rho(\lambda)$ is shown for the lattices with $10 \times 20 \times 10$ sites. It is seen in Fig.~\ref{fig:Dispersion_and_DOS}(d) that for Model~I with interactions between only the nearest sites taken into account, the DOS is symmetric with respect to $\lambda=0$, despite the absence of chiral symmetry in the model that requires a complete symmetry of the band structure with respect to $\lambda=0$. In turn, the introduction of additional couplings between the next-nearest (Model II) and the next-to-next nearest (Model III) sites breaks this DOS symmetry, as shown in Figs.~\ref{fig:Dispersion_and_DOS}(e) and~\ref{fig:Dispersion_and_DOS}(f).

Moreover, the DOS $\rho(\lambda)$ for Models~II and III in Figs.~\ref{fig:Dispersion_and_DOS}(e) and~\ref{fig:Dispersion_and_DOS}(f) features the presence of pronounced peaks at the upper bands that are also a hallmark of flat bands. Indeed, flat bands correspond to the vanishing group velocity $\partial \lambda/\partial k$ and therefore to a singularity in the density of states $\rho(\lambda) \propto \partial k/\partial \lambda$ for $\lambda \approx 5$.

Upon the introduction of bianisotropy, all three models feature the emergence of band gaps, as shown in Figs.~\ref{fig:Dispersion_and_DOS}(g)--~\ref{fig:Dispersion_and_DOS}(i) for $\Omega=7$. As discussed in the Supplemental Material~\cite{Supplement}, the width of the band gaps is linearly proportional to $\Omega$. In this case, all degeneracies are lifted, except for two-fold Kramers degeneracy. Moreover, the band structures for Models~II and III become more similar compared to the case without bianisotropy, as seen in Figs.~\ref{fig:Dispersion_and_DOS}(h) and~\ref{fig:Dispersion_and_DOS}(i). The formation of band gaps is also observed in the DOS shown in Figs.~\ref{fig:Dispersion_and_DOS}(j)--~\ref{fig:Dispersion_and_DOS}(l). While Model~I demonstrates the symmetry in DOS for non-zero values of the bianisotropy parameter, as shown in Fig.~\ref{fig:Dispersion_and_DOS}(j), the DOS for Models~II and III remain asymmetric with respect to $\lambda=0$; see Figs.~\ref{fig:Dispersion_and_DOS}(k) and~\ref{fig:Dispersion_and_DOS}(l). In contrast to the case with $\Omega=0$, there are no flat bands in the upper energy band for Models~II and III, as seen both from band diagrams and from the DOS. However, a nearly-flat band region is observed in the lower energy bands between the points $R$-$A$ in Figs.~\ref{fig:Dispersion_and_DOS}(h) and~\ref{fig:Dispersion_and_DOS}(i), with corresponding peaks in the DOS, Figs.~\ref{fig:Dispersion_and_DOS}(k) and~\ref{fig:Dispersion_and_DOS}(l). Finally, Models~II and III feature the emergence of in-gap eigenstates in DOS seen in Figs.~\ref{fig:Dispersion_and_DOS}(k) and~\ref{fig:Dispersion_and_DOS}(l).

\section{Eigenstates in finite models}
\label{sec:Finite_Model}

\subsection{Localization and spatial profiles of eigenstates}
\label{sec:Localization}

Next, we consider the spatial distributions of the eigenmodes and analyze their localization by evaluating the inverse participation ratios (IPR)~\cite{1974_Thouless}:
\begin{equation}
    \text{IPR}(\lambda) = \sum_{i,j,k}|\psi_{ijk}(\lambda)|^4,
    \label{eq:IPR}
\end{equation}
where $|\psi_{ijk}(\lambda)|=(\sum_{\alpha=1}^{4}|\psi_{ijk}^{\alpha}(\lambda)|^{2})^{1/2}$ is the amplitude of the eigenmode corresponding to the eigenvalue $\lambda$ at the site with coordinates ($i,j,k$) normalized by the condition $\sum_{i,j,k}|\psi_{ijk}|^{2} = 1$, and the index $\alpha$ denotes four on-site scalar components in the considered pseudospin basis $p_x + m_x$, $p_y + m_y$, $p_x - m_x$, and $p_y - m_y$. The summation is performed over all sites ($i,j,k$) in a finite tight-binding model. Such a definition allows us to characterize the localization of eigenstates simultaneously for all pseudospin components, in order to exclude the possible situation when a certain component is localized while some other one is not. For strongly delocalized states, IPR Eq.~\eqref{eq:IPR} takes low values of the order of $1/N$, where $N$ is the total number of sites, while for localized states it tends to unity (in the limiting case when the eigenmode is completely localized at one site).

\begin{figure}[tbp]
  \centering
  \includegraphics[width=8.5cm]{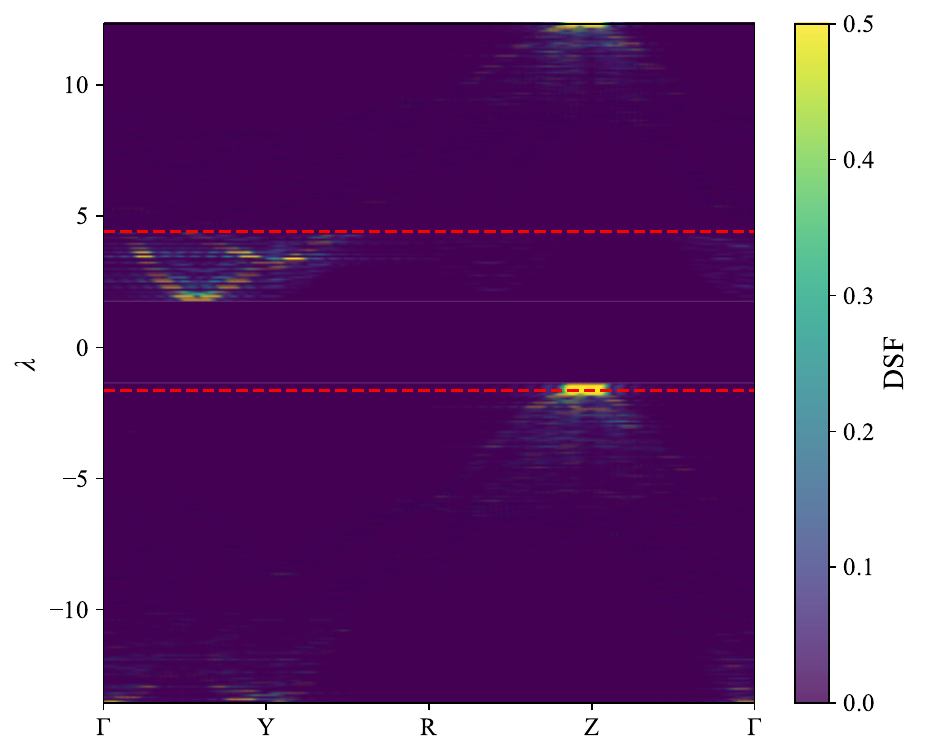}
  \caption{Dynamic structure factor $S(\lambda, k_{y}, k_{z})$ Eq.~\eqref{eq:DSF} for the Model~II with parameters of Fig.~\ref{fig:Localization} and the number of sites $20 \times 20 \times 20$ plotted along the trajectory $\Gamma$-$Y$-$R$-$Z$-$\Gamma$ in the Brillouin zone. Dashed horizontal lines denote the edges of the bulk band gap obtained as frequencies of the nearest bulk modes.}
  \label{fig:DSF}
\end{figure}

Figure~\ref{fig:Localization}(a) shows a numerically calculated spectrum of the tight-binding model corresponding to Model~II and consisting of two domains with bianisotropy parameters $\Omega = 7$ and $\Omega = -7$, each having the size $10 \times 10 \times 10$ sites. Thus, the resulting tight-binding model includes $20 \times 10 \times 10$ sites and features a domain wall in its middle. The points corresponding to the energy values in Fig.~\ref{fig:Localization}(a) are colored according to their IPR. It is seen that the bulk bands mostly include delocalized states with different numbers of wave nodes in the $x$-, $y$-, and $z$-directions, as shown in Fig.~\ref{fig:Localization}(b) and \ref{fig:Localization}(c). The states in the band gap are strongly localized at the domain wall, as seen in Fig.~\ref{fig:Localization}(d), \ref{fig:Localization}(e), and \ref{fig:Localization}(f). Such interface states also have different numbers of nodes in the $y$- and $z$-directions, with higher energy states having more nodes starting from zero in the lowest energy state in Fig.~\ref{fig:Localization}(d). Thus, the in-gap states seen in the DOS in Fig.~\ref{fig:Dispersion_and_DOS} represent interface states.

Along with delocalized states similar to Fig.~\ref{fig:Localization}(b) and \ref{fig:Localization}(c), the bulk bands demonstrate the presence of eigenmodes with higher localization, e.g., Fig.~\ref{fig:Localization}(g), \ref{fig:Localization}(h), and \ref{fig:Localization}(i). For example, there are eigenmodes in the form of hybridized interface states and bulk excitations, Fig.~\ref{fig:Localization}(g), and bulk modes with pronounced localization at the outer surfaces of domains; see Fig.~\ref{fig:Localization}(h). Finally, there are states localized at the domain wall with energies in the continuum that feature high IPR values, as demonstrated in Fig.~\ref{fig:Localization}(i). However, in contrast to in-gap states, such modes may hybridize with bulk excitations in a realistic system with losses taken into account, making their experimental implementation challenging.

\subsection{Dynamic structure factor of interface states}
\label{sec:DSF}

\begin{figure*}[t]
    \centering
    \includegraphics[width=17cm]{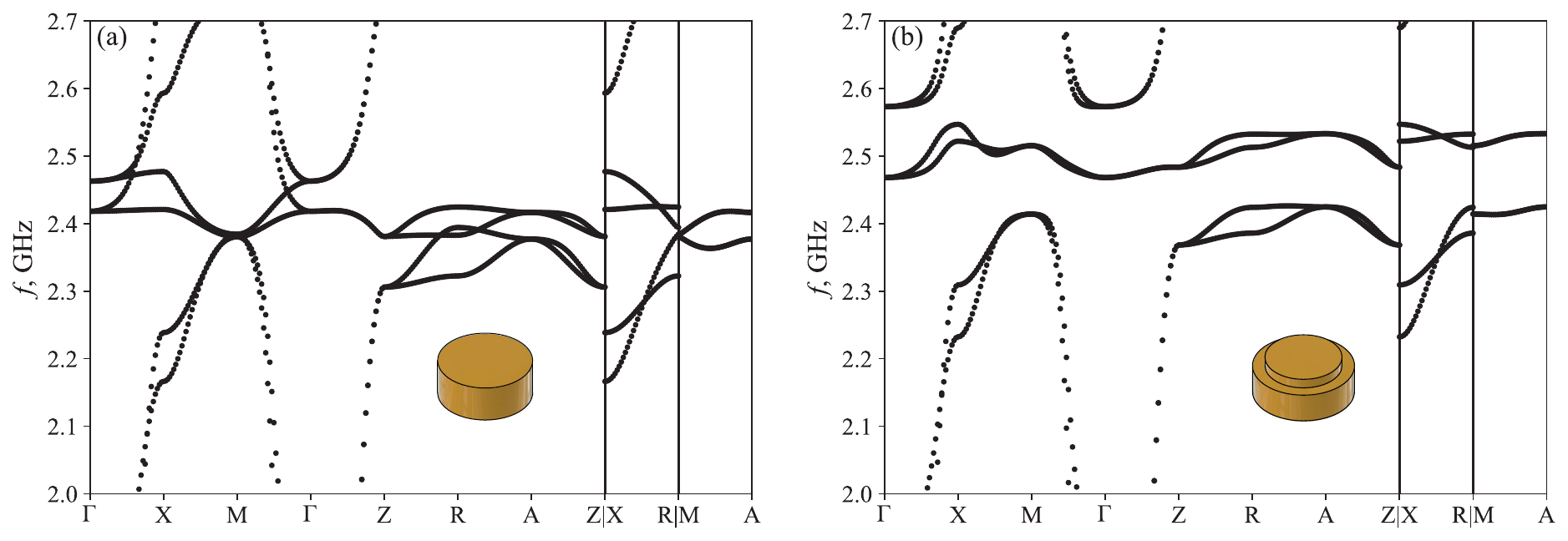}
    \caption{Numerically simulated bulk band diagrams for resonator arrays shown in Fig.~\ref{fig:System}(b) for (a) cylindrical resonators with height $h=12$~mm and diameter $d=29.1$~mm and (b) bianisotropic resonators composed of two cylinders with parameters $h_1=3$~mm, $d_1=22$~mm (the upper cylinder) and $h_2=9$~mm, $d_2=29.1$~mm (the lower cylinder). In both cases, the lattice constant in the $xy$ plane is $a_{x,y}=45.1$~mm and the lattice constant along the $z$-axis is $a_{z}=38$~mm. Insets demonstrate the geometry of the resonators.}
    \label{fig:CST}
\end{figure*}

To characterize the dispersion of interface states in the finite Model~II considered in Fig.~\ref{fig:Localization}, we evaluate dynamic structure factor (DSF) $S(\lambda,k_{y},k_{z})$
\begin{equation}
   S(\lambda,k_{y},k_{z}) = \sum_{n,l}\sum_{\alpha=1}^{4}\psi_{mnl}^{\alpha}(\lambda){\rm e}^{-(ik_{y}n+ik_{z}l)},
   \label{eq:DSF}
\end{equation}
where $\psi_{mnl}^{\alpha}(\lambda)$ is the $\alpha$-component of the eigenmode corresponding to the eigenvalue $\lambda$ with indices $-L_{x} \le m \le L_{x}$, $-L_{y} \le n \le L_{y}$, and $-L_{z} \le l \le L_{z}$ enumerating sites in the $x$-, $y$-, and $z$-directions in the system with sizes $2L_{x}$, $2L_{y}$, and $2L_{z}$, and $-\pi < k_{y}, k_{z} \le \pi$ are the wave numbers in the $y$- and $z$-directions, respectively. The value of index $m$ corresponds to the interface located in the $yz$ plane at $x=0$. Such a quantity represents Fourier transform of the eigenstates in the $yz$ plane, and, thus, its values indicate contributions from periodic waves with certain wave numbers $(k_{y},k_{z})$ in a given eigenstate with eigenvalue $\lambda$. In practice, well-resolved curves correspond to periodic oscillations, while elongated horizontal lines indicate localized states.

Figure~\ref{fig:DSF} demonstrates the DSF distribution along the trajectory between high-symmetry points $\Gamma(0,0,0)$-$Y(0,\pi,0)$-$R(0,\pi,\pi)$-$Z(0,0,\pi)$-$\Gamma(0,0,0)$ evaluated at the interface $x=0$ for the finite Model~II composed of two regions with $\Omega=-7$ and $\Omega=7$ having sizes $10 \times 20 \times 20$ sites in the $x$-, $y$-, and $z$-directions and forming an interface in the $yz$ plane, similarly to Fig.~\ref{fig:Localization}. It is seen that the DSF distribution is most pronounced in the band gap $-1.65 < \lambda < 4.43$, corresponding to the leading contribution of in-gap surface states. However, the contributions of bulk excitations with non-zero amplitudes at the interface are also observed at eigenvalues $\lambda<-1.65$ and $\lambda>4.43$ corresponding to bulk bands. As seen from Fig.~\ref{fig:DSF}, the region of surface states dispersion is concentrated between $\Gamma$-$Y$-$R$ points, while at other wave numbers the dispersion is less expressed. However, a broadening of the DSF distribution is observed associated with a relatively small size of the system. These size effects also result in a considerable separation between the adjacent eigenvalues leading to discrete horizontal lines of DSF values in Fig.~\ref{fig:DSF}. Finally, it is seen that the dispersion of surface states does not span the entire bulk band gap for the considered high value of $\Omega=\pm7$. The distributions of DSF for the finite Model~II with the bianisotropy parameter $\Omega=\pm 7$ and a different number of sites, as well as with $\Omega=\pm 5$ corresponding to a different width of the band gap, are discussed in the Supplemental Material.

\section{Comparison with numerical simulations}
\label{sec:Simulations}

To explore the accuracy and limitations of the introduced theoretical model and evaluate its potential experimental implementation, we compare band dispersions obtained within different tight-binding approximations in Section~\ref{sec:Dispersion} with full-wave numerical simulations performed in CST Microwave Studio. We consider a 3D lattice of cylindrical resonators depicted in Fig.~\ref{fig:System}(b). The permittivity of cylinders is $\varepsilon=39$ and the loss angle is $\delta=10^{-4}$ which correspond to $\text{MgO}-\text{CaO}-\text{TiO}_{2}$ ceramics. Such resonators have been applied in several experiments for 2D~\cite{2019_Slobozhanyuk, 2025_Rozenblit} and 3D~\cite{2026_Zhirihin} systems.

First, we start with the case of cylindrical resonators with height $h=12$~mm and diameter $d=29.1$~mm corresponding to the absence of bianisotropy, as shown in the inset of Fig.~\ref{fig:CST}(a). As seen from the band diagram in Fig.~\ref{fig:CST}(a), the bulk band gap closes with a quadratic degeneracy at the $M$ point, in agreement with the theoretical models in Fig.~\ref{fig:Dispersion_and_DOS}(a)-(c). However, several differences are observed. First, the numerically simulated band diagram features the presence of several additional bands that are not reproduced in the considered four-band theoretical models Eqs.~\eqref{eq:H_I_spin_basis_component}-\eqref{eq:H_III_spin_basis_component}. To describe these bands theoretically, one should take into account additional dipole modes (e.g., electric and magnetic $z$-dipoles) or higher-order multipoles. Second, the numerically simulated bands demonstrate a linear dispersion at the $\Gamma$ point in contrast to the tight-binding models that are obtained in the near-field approximation and do not take into account intermediate field terms and far-field radiation, which is essential for describing the light cone.

Next, we introduce bianisotropy by considering the resonators with broken inversion symmetry along the $z$-axis shown in the inset of Fig.~\ref{fig:CST}(b) that are composed of two cylinders with the following parameters: $h_1=3$~mm, $d_1=22$~mm (the upper cylinder) and $h_2=9$~mm, $d_2=29.1$~mm (the lower cylinder). The formation of a bulk band gap is observed, Fig.~\ref{fig:CST}(b), in full agreement with theoretical models in Fig.~\ref{fig:Dispersion_and_DOS}(g)-(i) and analogous results for 2D lattices composed of such resonators~\cite{2019_Slobozhanyuk, 2025_Rozenblit}.

It is seen that, despite their simplicity, the introduced tight-binding models successfully reproduce main properties of the numerically simulated systems. In particular, along with a quadratic degeneracy at the $M$ point, Models~II and III without bianisotropy successfully reproduce a wide loop formed by the bands along the $Z$-$R$-$A$ trajectory, their quadratic degeneracies at $Z$ and $A$ points, and a narrow loop along the $A$-$Z$ trajectory, as seen from the comparison of Fig.~\ref{fig:CST}(a) and Fig.~\ref{fig:Dispersion_and_DOS}(b),(c). However, along with a discrepancy at $\Gamma$ point associated with the light cone, the numerically simulated system demonstrates a lifted degeneracy of the modes along the $Z$-$R$-$A$-$Z$ trajectory that form two pairs of bands with similar properties. Between the $X$ and $R$ points, the bulk bands of theoretical Models~II and III and the numerically simulated system also demonstrate a good agreement (up to a lifted degeneracy in the numerical results): the presence of relatively flat upper bands and dispersive lower bands. In contrast to the $A$-$Z$ trajectory, the signs of group velocities for these bands are correctly reproduced in the theoretical models compared to the numerical results. The degeneracy of the bands between the $M$ and $A$ points observed in Fig.~\ref{fig:Dispersion_and_DOS}(b),(c) is also lifted in the numerical system. When bianisotropy is introduced, theoretical models in Fig.~\ref{fig:Dispersion_and_DOS}(h),(i) describe well the splitting of bands along the $M$-$A$ trajectory which is observed in the numerical simulations, Fig.~\ref{fig:CST}(b). However, while the band loops along the $Z$-$R$-$A$ and $A$-$Z$ trajectories are absent in the theoretical results in Fig.~\ref{fig:Dispersion_and_DOS}(h),(i), in numerical simulations they become less pronounced but are still present.

\section{Discussion}
\label{sec:Discussion}

In the present paper, we considered theoretically a tetragonal lattice of bianisotropic resonators applying the dyadic Green's function approach~\cite{2019_Gorlach,2025_Rozenblit} to extend the results of effective description based on perturbation theory~\cite{2017_Ochiai} and address the dispersion and localization properties of the eigenmodes. It is shown that the introduction of a bianisotropic response leads to the opening of a bulk band gap, similarly to the 2D hexagonal~\cite{2019_Slobozhanyuk} and square~\cite{2025_Rozenblit} resonator arrays, as well as 3D PTIs formed by stacking hexagonal layers~\cite{2017_Slobozhanyuk}.

In order to study the effects of long-range couplings that are known to be crucial for the description of a band gap formation in two-dimensional square lattices~\cite{2022_D4} or can even give rise to additional types of corner states in photonic HOTI~\cite{2020_Li}, we compared the band structures of three model Hamiltonians taking into account the couplings between the nearest, next-nearest, and next-to-next nearest sites. It is shown that a pronounced change in the band structure is observed when the next-nearest couplings are added, while the couplings in the third coordination sphere slightly modify the dispersion diagram. Moreover, the nearest-neighbor approximation is shown to result in additional degeneracies. Thus, it is necessary to take into account at least the next-nearest couplings in order to describe the properties of tetragonal lattices of bianisotropic resonators. The obtained results reproduce many characteristics of the numerically simulated band diagrams for an array of bianisotropic dielectric resonators, but demonstrate a significant discrepancy at $\Gamma$ point associated with the near-field approximation.

We propose the following directions for further development of the considered models. First, a detailed study of topological properties of the considered tetragonal arrays of bianisotropic resonators is of a principal interest in order to determine if these states are trivial or topological. As discussed in the Supplemental Material, the Berry curvature distributions in all three models zero out for $\Omega=0$, while upon the introduction of bianisotropy in Models~II and III, their Berry curvatures become non-vanishing, hinting at a potentially weak topological origin of the observed edge states. However, as can be directly checked, the corresponding spin Chern numbers are zeros. Moreover, the dispersion of interface states not always crosses the entire bulk band gap. While for a lower width of the band gap the interface states are gapless, as demonstrated in the Supplemental Material for $\Omega=\pm 5$, for a high band gap width they become gapped, as shown in Fig.~\ref{fig:DSF} for $\Omega=\pm 7$. This, in turn, points that the interface states in the considered models are likely to be trivial photonic Jackiw-Rebbi modes~\cite{2019_Gorlach} at the interface of regions with opposite signs of the effective mass $\Omega$. Compared to bianisotropic photonic structures of Fig.~\ref{fig:System}(b) and Fig.~\ref{fig:CST} that explicitly break the inversion symmetry along the $z$-axis, the considered theoretical models have different symmetry properties and demonstrate a pronounced difference in the band structure at $\Gamma$ point, which is related to the near-field approximation. The corresponding modification of the symmetry of the models and the incorporation of far-field terms in dyadic Green's functions may render important for the correct description of topological properties.

Second, it is seen that the considered models demonstrate the states localized at the domain wall with energies in the continuum of bulk bands, as shown in Fig.~\ref{fig:Localization}(i). It appears intriguing to study if these states correspond to bound states in the continuum~\cite{2020_Cerjan} or some different highly-localized modes~\cite{2022_D4} by introducing losses in the models to define quality factors of the modes. However, such a modification will render the Hamiltonians non-Hermitian, even if only ohmic losses are considered, and result in a non-linear eigenvalue problem if radiative losses are taken into account by considering complete dyadic Green's functions with intermediate- and far-field contributions. This may increase the complexity of the model, but can result in a set of interesting effects. There are examples of PTIs that demonstrate an interplay of near- and long-range couplings~\cite{2022_Fan}, and far-field properties of PTIs based on resonator arrays have been considered~\cite{2018_Gorlach, 2026_Zhirihin}.

Finally, the considered models include only electric and magnetic dipoles in the $(xy)$ plane. Taking into account dipoles along the $z$-axis will allow one to consider a more general case of the resonator shape when the approximations valid for the previous experimental realizations~\cite{2019_Slobozhanyuk, 2025_Rozenblit, 2026_Zhirihin} are not satisfied. Moreover, including a quadrupole near-field response~\cite{2022_Wu} or taking into account higher multipoles~\cite{2024_Mazanov} can also result in interesting physics. The models can be extended to more complex unit cells, e.g., by including dimerization or staggering~\cite{2021_Bobylev,2024_Sang} to realize a hierarchy of lower-dimensional hinge or corner states, or by considering nonlinear effects~\cite{2020_Smirnova}.

For experimental realization of the considered models at microwave frequencies, ceramic cylindrical resonators can be applied, in analogy to 2D hexagonal~\cite{2019_Slobozhanyuk} and square~\cite{2025_Rozenblit} lattices, as well as the recent demonstration of a 3D array of stacked hexagonal layers~\cite{2026_Zhirihin}. Moreover, metallic resonators implemented as printed circuit boards can also be considered~\cite{2016_Slobozhanyuk}. From a practical perspective, such resonator arrays can be applied to control field localization or implement signal routing along complex trajectories by designing specific shapes of the domain interfaces.

\begin{acknowledgments}

We acknowledge fruitful discussions with Dmitry Zhirihin and Georgy Kurganov. The work is supported by the Russian Science Foundation (project 24-79-10314).

\end{acknowledgments}


%

\end{document}


\title{Supplemental Material\\~\\Theoretical description of interface states in a tetragonal lattice of bianisotropic resonators}

\author{Alina D. Rozenblit}
\email{alina.rozenblit@metalab.ifmo.ru}
\affiliation{School of Physics and Engineering, ITMO University, 197101 Saint Petersburg, Russia}

\author{Nikita A. Olekhno}
\affiliation{School of Physics and Engineering, ITMO University, 197101 Saint Petersburg, Russia}

\date{\today}

\maketitle

\tableofcontents

\section{Bloch Hamiltonian}
\label{sec:Hamiltonian}

In this Section, we provide a detailed derivation of the Bloch Hamiltonian. We consider a cubic lattice with period $a$ formed by point electric ${\mathbf{p} = (p_{x}, p_{y})^{\rm T}}$ and magnetic $\mathbf{m} = (m_{x}, m_{y})^{\rm T}$ dipoles oriented in the $xy$-plane, as shown in Fig.~1(a) in the main text. The components of the excited electric $\mathbf{E} = (E_{x}, E_{y})^{\rm T}$ and magnetic $\mathbf{H} = (H_{x}, H_{y})^{\rm T}$ fields and induced dipole moments in a lattice site with coordinates $(ia, ja, ka)$ are related by the polarizability tensor $\widehat{\alpha}$ as
\begin{equation}
    \begin{pmatrix}
    \mathbf{p}^{ijk}\\
    \mathbf{m}^{ijk}
    \end{pmatrix}
    = 
    \widehat{\alpha}
    \begin{pmatrix}
    \mathbf{E}^{ijk}\\
    \mathbf{H}^{ijk}
    \end{pmatrix}.
    \label{eq:Dipole_moments}
\end{equation}
In turn, the polarizability tensor includes the electric $\widehat{\alpha}^{\rm ee}$, magnetic $\widehat{\alpha}^{\rm mm}$, electromagnetic $\widehat{\alpha}^{\rm em}$, and magnetoelectric $\widehat{\alpha}^{\rm me}$ components:
\begin{equation}
    \widehat{\alpha} =
    \begin{pmatrix}
    \widehat{\alpha}^{\rm ee} & \widehat{\alpha}^{\rm em}\\
    \widehat{\alpha}^{\rm me} & \widehat{\alpha}^{\rm mm}
    \end{pmatrix}.
\end{equation}

Next, we consider the case with the electromagnetic duality ($\widehat{\varepsilon} = \widehat{\mu}$) satisfied, resulting in equal electric and magnetic polarizabilities $\beta$ along the $x$- and $y$-axes, which yields
\begin{equation}
    \widehat{\alpha}^{\rm ee} = \widehat{\alpha}^{\rm mm} = \beta \cdot\widehat{\sigma}_0,
\end{equation}
where $\widehat{\sigma}_0 = (1,0;0,1)$ is the unity matrix. At the same time, reciprocal bianisotropic particles demonstrate the presence of non-vanishing electromagnetic (magnetoelectric) components described by the electromagnetic coupling $\chi$~\cite{2018_Asadchy}:
\begin{equation}
    \widehat{\alpha}^{\rm em} = -\{\widehat{\alpha}^{\rm me}\}^{\rm T} = -\chi \cdot \widehat{\sigma}_2,
\end{equation}
where $\widehat{\sigma}_2 = (0, -{\rm i}; {\rm i}, 0)$ is the Pauli matrix. Thus, the polarizability tensor takes the following form:
\begin{equation}
    \widehat{\alpha} =
    \begin{pmatrix}
        \beta & 0 & 0 & {\rm i} \chi\\
        0 & \beta & -{\rm i} \chi & 0\\
        0 & {\rm i} \chi & \beta & 0\\
        -{\rm i} \chi & 0 & 0 & \beta
    \end{pmatrix}.
\end{equation}
The electromagnetic field components can be expressed through dipole moments with the help of the inverse polarizability tensor $\widehat{\alpha}^{-1}$:
\begin{equation}
    \begin{pmatrix}
    \mathbf{E}^{ijk}\\
    \mathbf{H}^{ijk}
    \end{pmatrix} =
    \widehat{\alpha}^{-1}
    \begin{pmatrix}
    \mathbf{p}^{ijk}\\
    \mathbf{m}^{ijk}
    \end{pmatrix},
    \label{eq:Fields_by_dipole_moments}
\end{equation}
where
\begin{equation}
    \widehat{\alpha}^{-1}
    = \dfrac{1}{\beta^2-\chi^2}
    \begin{pmatrix}
        \beta & 0 & 0 & -{\rm i}\chi\\
        0 & \beta & {\rm i}\chi & 0\\
        0 & -{\rm i}\chi & \beta & 0\\
        {\rm i}\chi & 0 & 0 & \beta
    \end{pmatrix}.
    \label{Inverse_polarizability_tensor}
\end{equation}

For simplicity, we assume that the diagonal elements of $\widehat{\alpha}^{-1}$ depend on the resonance frequency $f_0$ as ${\beta/(\beta^2-\chi^2) \propto f - f_0}$, while the term $\chi/(\beta^2-\chi^2)$ can be approximated as a constant~\cite{2023_Kim}. Converting $\widehat{\alpha}^{-1}$ to dimensionless values and taking into account the approximations, we obtain eigenvalues {$\lambda = a^3 \beta / (\beta^2-\chi^2)$} and the bianisotropy parameter {$\Omega = a^3 \chi / (\beta^2-\chi^2)$}. Thus,~\eqref{eq:Fields_by_dipole_moments} can be rewritten in the following form:
\begin{equation}
    \begin{pmatrix}
    \mathbf{E}^{ijk}\\
    \mathbf{H}^{ijk}
    \end{pmatrix} =
    \Big(\dfrac{\lambda}{a^3} \widehat{\sigma}_0 \otimes \widehat{\sigma}_0 + \dfrac{\Omega}{a^3} \widehat{\sigma}_1 \otimes \widehat{\sigma}_2\Big)
    \begin{pmatrix}
        \mathbf{p}^{ijk}\\
        \mathbf{m}^{ijk}
    \end{pmatrix},
    \label{eq:Inverse_polarizability_approximated}
\end{equation}
where $\widehat{\sigma}_1 = (0,1;1,0)$ is the Pauli matrix.

On the other hand, the electromagnetic field amplitudes at a certain site can be found as the sum of fields excited by dipoles from other sites and expressed via dyadic Green functions:
\begin{equation}
    \begin{pmatrix}
        \mathbf{E}^{ijk} \\
        \mathbf{H}^{ijk}
    \end{pmatrix} = \sum_{\substack{m \ne i\\
    n \ne j \\ l \ne k}}
    \begin{pmatrix}
        \widehat{G}^{\rm ee}(r, k_0) & \widehat{G}^{\rm em}(r, k_0)\\
        \widehat{G}^{\rm me}(r, k_0) & \widehat{G}^{\rm mm}(r, k_0)
    \end{pmatrix}
        \begin{pmatrix}
        \mathbf{p}^{mnl} \\
        \mathbf{m}^{mnl}
    \end{pmatrix}.
    \label{eq:E_and_Green}
\end{equation}
where $r = a\sqrt{(i-m)^2+(j-n)^2+(k-l)^2}$ is the distance between the lattice sites with coordinates $(ma,na,la)$ and $(ia,ja,ka)$, and $k_0$ is the wave number. The bianisotropic parts of the dyadic Green function are related as $\widehat{G}^{\rm em}=-\widehat{G}^{\rm me}$, while the electromagnetic duality requires $\widehat{G}^{\rm ee}=\widehat{G}^{\rm mm}$. Dyadic Green functions can be calculated using the following equations~\cite{1994_Tai}:
\begin{align}
    G^{\rm ee}_{\zeta\eta} & = (\partial_{\zeta} \partial_{\eta} + k^2_0\delta_{\zeta\eta})\dfrac{e^{{\rm i} k_0 r_{ij}}}{r_{ij}},
    \label{eq:Gee} \\
    G^{\rm em}_{\zeta\eta} & = {\rm i} k_0 \varepsilon_{\zeta\eta\kappa}\partial_{\kappa}\dfrac{e^{{\rm i} k_0 r_{ij}}}{r_{ij}},
    \label{eq:Gem}
\end{align}
where $\delta_{\zeta\eta}$ stands for the Kronecker delta, $\varepsilon_{\zeta \eta \kappa}$ is the Levi-Civita symbol, and $\partial_\zeta = \partial/\partial\zeta$. The parameters $\zeta$, $\eta$, $\kappa$ take the values $\{x,y,z\}$. Considering the spacing between the dipoles as a function of the distance along the $x$-, $y$-, and $z$-axes {$r = \sqrt{x^2+y^2+z^2}$}, we obtain the following expressions for the dyadic Green function components:
\begin{align}
     G^{\rm ee}_{xx} & = \dfrac{e^{{\rm i} k_0 r}}{r^3}\Big(\dfrac{x^2}{r^2}(3-3 {\rm i} k_0 r - k_0^2 r^2)+{\rm i} k_0 r + k^2_0 r^2 - 1\Big), \nonumber\\
    G^{\rm ee}_{yy} & = \dfrac{e^{{\rm i} k_0 r}}{r^3}\Big(\dfrac{y^2}{r^2}(3-3 {\rm i} k_0 r - k_0^2 r^2)+{\rm i} k_0 r + k^2_0 r^2 - 1\Big), \nonumber\\
    G^{\rm ee}_{xy} & = G^{\rm ee}_{yx}= \dfrac{e^{{\rm i} k_0 r}}{r^3}\Big(\dfrac{xy}{r^2}(3-3 {\rm i} k_0 r - k_0^2 r^2)\Big), \nonumber\\
    G^{\rm em}_{xx} & = -G^{\rm em}_{yy}= 0, \nonumber\\
    G^{\rm em}_{xy} & = -G^{\rm em}_{yx}= \dfrac{e^{{\rm i} k_0 r}}{r^3}(-{\rm i} k_{0} r -k_{0}^{2}r^2).
    \label{G_general}
\end{align}

In the following, we focus on the near-field approximation ($k_{0}r \ll 1$) of the dyadic Green function, in which only the near-field terms proportional to $1/r^3$ remain, and rewrite the distances between the sites in~\eqref{G_general} through their coordinates:
\begin{align}
    G^{\rm ee}_{xx} & = \dfrac{3a^2(m-i)^2}{r^5}- \dfrac{1}{r^3}, \nonumber\\
    G^{\rm ee}_{yy} & = \dfrac{3a^2(n-j)^2}{r^5}- \dfrac{1}{r^3}, \nonumber\\
    G^{\rm ee}_{xy} & = G^{\rm ee}_{yx}= \dfrac{3a^2(m-i)(n-j)}{r^5}, \nonumber\\
    G^{\rm em}_{xx} & = G^{\rm em}_{yy} =G^{\rm em}_{xy} = G^{\rm em}_{yx} = 0.
    \label{eq:G_near_field}
\end{align}

As the first model, we consider the cubic lattice with interactions only between the nearest sites ($r=a$). In this case,~\eqref{eq:E_and_Green} has the following form:
\begin{multline}
    \begin{pmatrix}
        \mathbf{E}^{ijk} \\
        \mathbf{H}^{ijk}
    \end{pmatrix} = \sum_{\substack{m = i \pm 1\\
    n = j \\ l = k}}
    \begin{pmatrix}
        \widehat{G}^{\rm ee} & \widehat{G}^{\rm em}\\
        \widehat{G}^{\rm me} & \widehat{G}^{\rm mm}
    \end{pmatrix}
        \begin{pmatrix}
        \mathbf{p}^{mnl} \\
        \mathbf{m}^{mnl}
    \end{pmatrix} + 
    \sum_{\substack{m = i \\
    n = j \pm 1 \\ l = k}}
    \begin{pmatrix}
        \widehat{G}^{\rm ee} & \widehat{G}^{\rm em}\\
        \widehat{G}^{\rm me} & \widehat{G}^{\rm mm}
    \end{pmatrix}
        \begin{pmatrix}
        \mathbf{p}^{mnl} \\
        \mathbf{m}^{mnl}
    \end{pmatrix}
    +
    \sum_{\substack{m = i\\
    n = j \\ l = k \pm 1}}
    \begin{pmatrix}
        \widehat{G}^{\rm ee} & \widehat{G}^{\rm em}\\
        \widehat{G}^{\rm me} & \widehat{G}^{\rm mm}
    \end{pmatrix}
        \begin{pmatrix}
        \mathbf{p}^{mnl} \\
        \mathbf{m}^{mnl}
    \end{pmatrix}.
    \label{eq:E_and_Green_I_general}
\end{multline}
Applying Bloch's theorem results in~\eqref{eq:E_and_Green_I_general} taking the following form:
\begin{equation}
    \begin{pmatrix}
        \mathbf{E}^{ijk}\\
        \mathbf{H}^{ijk}
    \end{pmatrix} = a^{-3}\Big((\cos{k_{x}a} + \cos{k_{y}a} - 2\cos{k_{z}a})\widehat{\sigma}_0 \otimes \widehat{\sigma}_0 + 3 (\cos{k_{x}a} - \cos{k_{y}a})\widehat{\sigma}_0 \otimes \widehat{\sigma}_3 \Big)
    \begin{pmatrix}
        \mathbf{p}^{ijk}\\
        \mathbf{m}^{ijk}
    \end{pmatrix}.
    \label{eq:E_and_G_prefinal}
\end{equation}

Finally, we compare the right hand sides of~\eqref{eq:Inverse_polarizability_approximated} and~\eqref{eq:E_and_G_prefinal}, consider dimensionless units $a=1$, and rewrite the result in the form of an eigenvalue problem $\widehat{H}^{\prime}_{\rm I}\ket{\psi^{\prime}}=\lambda\ket{\psi^{\prime}}$. Thus, the effective Bloch Hamiltonian in the basis $\ket{\psi^{\prime}} = (p_x, p_y, m_x, m_y)^{\rm T}$ in the nearest-neighbor approximation (denoted by the lower index~I) is given by the following expression:
\begin{equation}
    \widehat{H}^{\prime}_{\rm I} = (\cos{k_x}+\cos{k_y}-2\cos{k_z}) \widehat{\sigma}_0 \otimes \widehat{\sigma}_0 + 3(\cos{k_x}-\cos{k_y}) \widehat{\sigma}_0 \otimes \widehat{\sigma}_3 - \Omega \widehat{\sigma}_1 \otimes \widehat{\sigma}_2
 \label{eq:H_I_old_basis}.
\end{equation}

To convert the obtained Bloch Hamiltonian $\widehat{H}^{\prime}_{\rm I}$ to a block-diagonal form, the basis $\ket{\psi^{\prime}}$ should be replaced by the pseudospin basis $\ket{\psi}=(p_x + m_x, p_y + m_y, p_x - m_x, p_y - m_y)^{\rm T}$~\cite{2017_Slobozhanyuk}. Thus, we apply the following unitary transformation matrix $\widehat{U}$:
\begin{equation}
    \widehat{U} = \dfrac{1}{\sqrt{2}}
    \begin{pmatrix}
        1 & 0 & 1 & 0\\
        0 & 1 & 0 & 1\\
        1 & 0 & -1 & 0\\
        0 & 1 & 0 & -1
    \end{pmatrix},
\end{equation}
and evaluate $\widehat{H}_{\rm I} = \widehat{U}\widehat{H}^{\prime}_{\rm I}\widehat{U}^{\dag}$. Finally, the full Bloch Hamiltonian $\widehat{H}$ in the pseudospin basis has the form
\begin{equation}
    \widehat{H} = 
    \begin{pmatrix}
        \widehat{H}^{\uparrow} & 0\\
        0 & \widehat{H}^{\downarrow}
    \end{pmatrix},
 \label{eq:H_I_spin_basis_full}
\end{equation}
where $\widehat{H}^{\uparrow(\downarrow)}$ are Bloch Hamiltonians for two pseudospins denoted as $\uparrow(\downarrow)$, respectively. Within the nearest neighbor approximation (Model~I), these pseudospin Hamiltonians are given by
\begin{equation}
    \widehat{H}^{\uparrow(\downarrow)}_{\rm I} =
    (\cos{k_x}+\cos{k_y}-2\cos{k_z}) \widehat{\sigma}_0 + 3(\cos{k_x}-\cos{k_y}) \widehat{\sigma}_3 \pm \Omega \widehat{\sigma}_2.
    \label{eq:H_I_spin_basis_component}
\end{equation}

Next, we consider the model that includes the couplings between the nearest ($r=a$) and the next-nearest ($r = \sqrt{2}a$) sites (Model~II), as well as the model that takes into account the interactions between the sites in the first ($r=a$), the second ($r = \sqrt{2}a$), and the third ($r=\sqrt{3}a$) coordination spheres (Model~III) in~\eqref{eq:E_and_Green_I_general}. Then, we obtain the following Bloch Hamiltonians $\widehat{H}^{\uparrow(\downarrow)}_{\rm II}$ and $\widehat{H}^{\uparrow(\downarrow)}_{\rm III}$ in the pseudospin basis $\ket{\psi}=(p_x + m_x, p_y + m_y, p_x - m_x, p_y - m_y)^{\rm T}$:
\begin{multline}
    \widehat{H}_{\rm II}^{\uparrow(\downarrow)} = \Big(\cos{k_x}+\cos{k_y}-2\cos{k_z} + \dfrac{2\cos{k_x} \cos{k_y} - \cos{k_x} \cos{k_z} - \cos{k_y} \cos{k_z}}{2\sqrt{2}}\Big) \widehat{\sigma}_0  + \\ + 3\Big(\cos{k_x}-\cos{k_y} + \dfrac{\cos{k_x} \cos{k_z}-\cos{k_y} \cos{k_z}}{2\sqrt{2}}\Big)  \widehat{\sigma}_3 - \dfrac{3}{\sqrt{2}} \sin{k_x} \sin{k_y} \widehat{\sigma}_1 \pm \Omega \widehat{\sigma}_2
    \label{eq:H_II_spin_basis_component}
\end{multline}
and 
\begin{multline}
    \widehat{H}_{\rm III}^{\uparrow(\downarrow)}
    = \Big(\cos{k_x}+\cos{k_y}-2\cos{k_z} + \dfrac{2\cos{k_x} \cos{k_y} - \cos{k_x} \cos{k_z} - \cos{k_y} \cos{k_z}}{2\sqrt{2}}\Big) \widehat{\sigma}_0 + \\ + 3\Big(\cos{k_x}-\cos{k_y} + \dfrac{\cos{k_x} \cos{k_z}-\cos{k_y} \cos{k_z}}{2\sqrt{2}}\Big) \widehat{\sigma}_3 - \Big(\dfrac{3}{\sqrt{2}}+\dfrac{8}{3\sqrt{3}}\cos{k_z}\Big) \sin{k_x} \sin{k_y} \widehat{\sigma}_1 \pm \Omega \widehat{\sigma}_2.
    \label{eq:H_III_spin_basis_component}
\end{multline}

\section{Band dispersion}
\label{sec:Eigenvalues}

\begin{figure*}[b]
    \centering
    \includegraphics[width=8cm]{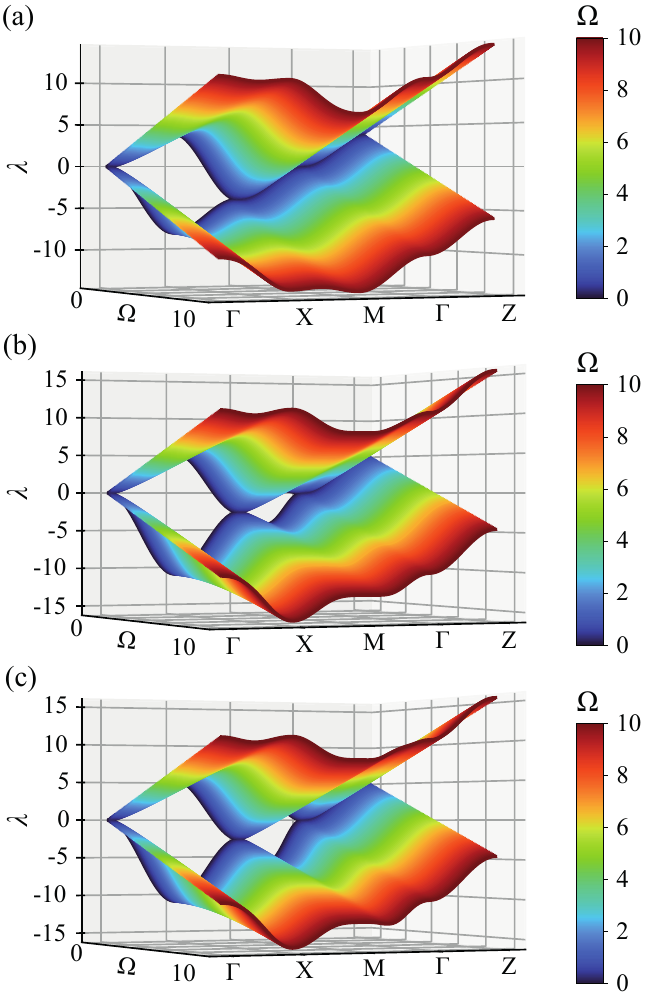}
    \caption{The dispersion diagrams for different values of the bianisotropy parameter $\Omega$ along $\Gamma$-$X$-$M$-$\Gamma$-$Z$ trajectory between the high-symmetry points of the Brillouin zone for (a) Model~I,~\eqref{eq:H_I_spin_basis_component}; (b) Model~II,~\eqref{eq:H_II_spin_basis_component}; and (c) Model~III,~\eqref{eq:H_III_spin_basis_component}}.
    \label{fig:Dispersion}
\end{figure*}

Analytical expressions for the energy bands of pseudospin Hamiltonians~\eqref{eq:H_II_spin_basis_component} and~\eqref{eq:H_III_spin_basis_component} have the following form. For Model~I,
\begin{equation}
    \lambda_{\rm I} = \cos{k_x} + \cos{k_y} - 2\cos{k_z} \pm 3\Bigl\{1+\dfrac{\Omega^2}{9}+\dfrac{\cos{(2 k_x)} + \cos{(2k_y)}}{2} - 2\cos{k_x}\cos{k_y} \Bigr\}^{1/2}.
    \label{eg:lambda_I}
\end{equation}
For Model~II, the eigenvalues are
\begin{multline}
   \lambda_{\rm II} = \cos{k_x}+\cos{k_y}-2\cos{k_z}+\dfrac{1}{2\sqrt{2}} \bigl (2\cos{k_x}\cos{k_y}-\cos{k_x}\cos{k_z}-\cos{k_y}\cos{k_z} \bigr) \pm \\
   \pm \dfrac{1}{8}\Bigl\{684 + 64\Omega^2+234\cos{(2k_x)}-1224\cos{k_x}\cos{k_y}+234\cos{(2k_y)} + \\
     + 72\cos{(2k_x)}\cos{(2k_y)}+36 \bigl (\cos{k_x}-\cos{k_y} \bigr)^2 \bigl (8\sqrt{2}\cos{k_z}
    +\cos{(2k_z)} \bigr)\Bigr\}^{1/2}.
    \label{eg:lambda_II}
\end{multline}
Finally, for Model~III,
\begin{multline}
    \lambda_{\rm III} = \cos{k_x}+\cos{k_y}-2\cos{k_z} +\dfrac{1}{2\sqrt{2}}(2\cos{k_x}\cos{k_y}-\cos{k_x}\cos{k_z}-\cos{k_y}\cos{k_z}) \pm \\
    \pm \dfrac{1}{12\sqrt{6}}\Bigl\{9490 + 864\Omega^2 +2903\cos{(2k_y)}-72\sqrt{2} \bigl (-54-8\sqrt{3}+(8\sqrt{3}-27)\cos{(2k_y)} \bigr)\cos{k_z} + \\
    + \bigl ( 742 -13\cos{(2k_y)} \bigl) \cos{(2k_z)} - 972\cos{k_x}\cos{k_y} \bigl (17 + 8\sqrt{2}\cos{k_z} + \cos{(2k_z)} \bigr ) + \\
    + \cos{(2k_x)} \Bigl ( 2903+72\sqrt{2}(27-8\sqrt{3})\cos{k_z} - 13\cos{(2k_z)} +\\
    + 4\cos{(2k_y)} \bigl (307 + 144\sqrt{6}\cos{k_z} + 64\cos{(2k_z)} \bigr) \Bigr)\Bigr\}^{1/2}.
    \label{eg:lambda_III}
\end{multline}
As seen from~\eqref{eg:lambda_I}--~\eqref{eg:lambda_III}, the eigenvalues are linearly proportional to the value of the bianisotropy parameter $\lambda \propto |\Omega|$, which is observed in Fig.~\ref{fig:Dispersion}.

\section{Berry curvature distributions}
\label{sec:Berry_curvature}

\begin{figure*}[b]
    \centering
    \includegraphics[width=16cm]{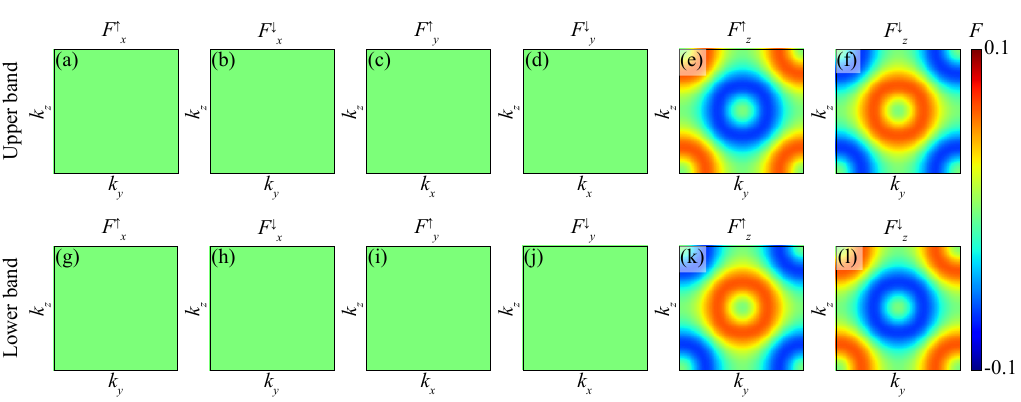}
    \caption{Numerically evaluated Berry curvature distributions for two pseudospins for the (a)-(f) upper and (g)-(l) lower energy bands of the Hamiltonian Eq.~\eqref{eq:H_II_spin_basis_component} with bianisotropy parameter $\Omega = 7$.}
    \label{fig:Berry_Curvature_Model_II}
\end{figure*}

To study the topological properties of the considered models, we visualize the distributions of Berry curvature $F_{\xi}^{\uparrow(\downarrow)}$ calculated from the normalized Bloch Hamiltonian eigenvectors $\psi(k)$ for three orthogonal planes of the cubic lattice using the following expression~\cite{2004_Haldane}:
\begin{equation}
    F_{\xi}^{\uparrow(\downarrow)}(k_{\zeta}, k_{\eta}, k_{\xi}=0) = \dfrac{\partial}{\partial k_{\zeta}}\braket{\psi_{1(2)}^{\uparrow(\downarrow)} | \dfrac{\partial}{\partial k_{\eta}}| \psi_{1(2)}^{\uparrow(\downarrow)}} - \dfrac{\partial}{\partial k_{\eta}}\braket{\psi_{1(2)}^{\uparrow(\downarrow)} | \dfrac{\partial}{\partial k_{\zeta}}| \psi_{1(2)}^{\uparrow(\downarrow)}},
    \label{eq:Berry_Curvature}
\end{equation}
where the indices $\zeta \ne \eta \ne \xi$ take the values $\{x, y, z\}$, the subscript $1(2)$ denotes the upper or lower eigenvalue band of the pseudospin Hamiltonians, and the superscript $\uparrow$($\downarrow$) corresponds to the pseudospin-up or pseudospin-down Hamiltonians $\widehat{H}^{\uparrow(\downarrow)}$, respectively.

\begin{figure*}[t]
    \centering
    \includegraphics[width=16cm]{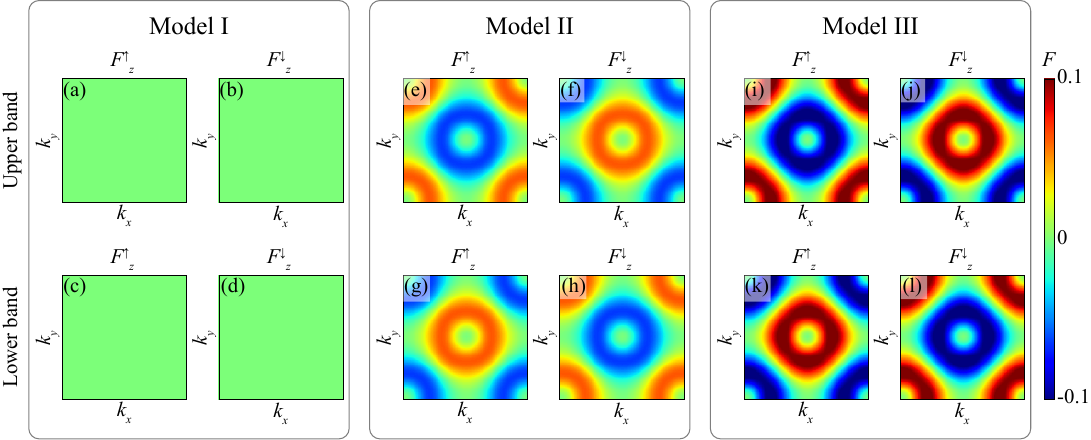}
    \caption{Numerically calculated distributions of the Berry curvature $F_z$ for (a)-(f) Model~I, (e)-(h) Model~II, and (i)-(l) Model~III for the two pseudospins and two energy bands in the case of bianisotropy parameter $\Omega = 7$.}
    \label{fig:Berry_curature}
\end{figure*}

We proceed with the calculation of Berry curvatures for Model~II. In the absence of bianisotropy ($\Omega = 0$), each of the eigenvalue bands in the considered model features vanishing Berry curvatures $F_{x}^{\uparrow(\downarrow)}$, $F_{y}^{\uparrow(\downarrow)}$, and $F_{z}^{\uparrow(\downarrow)}$ in the Brillouin zone, pointing to the trivial topology~\cite{2020_Kim_Recent}. Upon the introduction of non-zero bianisotropy parameter $\Omega = 7$, a non-zero $F_z$ distribution emerges in the $xy$-plane, as shown in Fig.~\ref{fig:Berry_Curvature_Model_II}. In particular, two regions are observed around $M$ and $\Gamma$ high symmetry points with opposite signs of the Berry curvature $F_z$. Since $F_x$ and $F_y$ zero out even in the presence of bianisotropic coupling, the considered tetragonal resonator array may represent either a trivial structure or a weak PTI formed by the stacking of 2D square lattice layers~\cite{2025_Rozenblit} along the $z$ direction, in analogy to a 3D structure of stacked hexagonal layers~\cite{2026_Zhirihin}. It is seen that Berry curvature distributions change sign for different band indices and opposite pseudospins.

Next, we compare the Berry curvatures $F_z(k_x, k_y)$ for each of the considered approximations in the case of a nonzero bianisotropy parameter $\Omega = 7$. The upper and lower bands of the pseudospin-up and pseudospin-down Hamiltonians~\eqref{eq:H_I_spin_basis_component}, corresponding to Model~I, possess the vanishing Berry curvature $F_z(k_x, k_y)$ for any values of wavenumbers $k_x$ and $k_y$, as shown in Figs.~\ref{fig:Berry_curature}(a)--\ref{fig:Berry_curature}(d). The results obtained for Model~I agree with the absence of in-gap eigenstates in DOS seen in Fig.~2(j) in the main text and demonstrate that in order to describe the interface states in such lattices of bianisotropic resonators, it is necessary to take into account the interactions between the resonators within at least the second coordination sphere. However, Berry curvature $F_z$ is no longer zero as soon as the couplings between the next-nearest sites are added to the model, as illustrated in Figs.~\ref{fig:Berry_curature}(e)--\ref{fig:Berry_curature}(h). In particular, two regions are observed around $\Gamma$ and $M$ high-symmetry points with opposite signs of Berry curvature $F_z$. The change between the bands and pseudospins is accompanied by a change in the sign of Berry curvature. The introduction of the interactions between the lattice sites in the third coordination sphere demonstrates the features of the Berry curvature distributions similar to those obtained for Model~II. The main difference is the increase in non-zero values of the Berry curvature, as depicted in Figs.~\ref{fig:Berry_curature}(i)--\ref{fig:Berry_curature}(l). However, the integral over the Brillouin zone, i.e., spin Chern number, is zero for all $F_{z}$ distributions in the considered models.

\section{Model with a double-bent domain wall}
\label{sec:Zigzag}

\begin{figure*}[t]
    \centering
    \includegraphics[width=16cm]{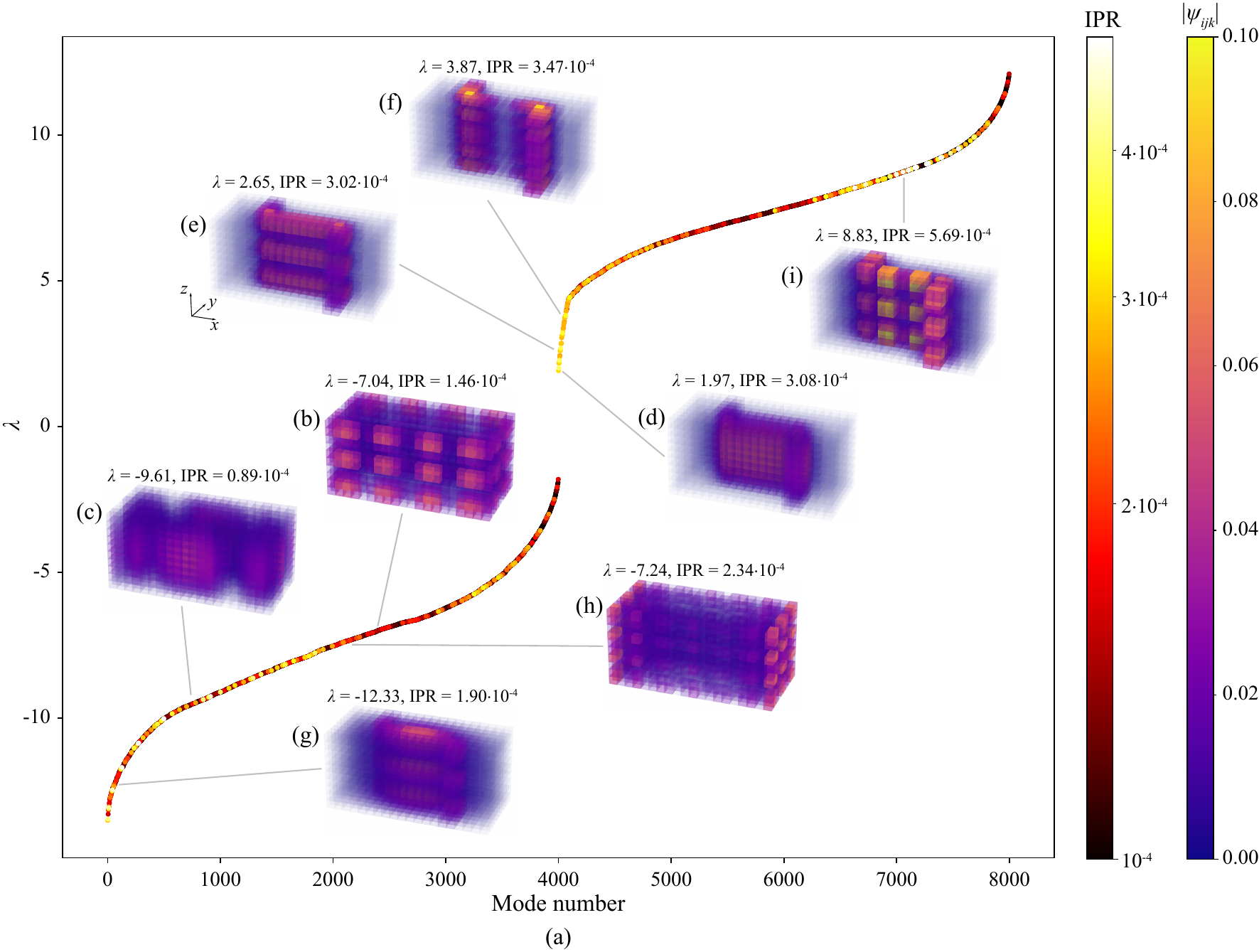}
    \caption{(a) Spectrum of eigenvalues $\lambda$ for two domain the system with double bent domain wall and bianisotropy parameters $\Omega = 7$ and $\Omega = -7$, respectively. Color shows the inverse participation ratio ($\rm IPR$). (b)-(i) Eigenfunctions profiles corresponding to the absolute value of pseudospin-up polarization $p_x + m_x$ shown by color, which demonstrate (b),(c) bulk states, (d)-(f) interface states localized at the domain wall in the bandgap, (g) an interface state hybridized with a bulk mode, (h) a bulk mode with dominant surface localization at the boundary, and (i) an edge state in the continuum.}
    \label{fig:Localization_zigzag}
\end{figure*}

To study the dependence of the interface states on its geometry, we consider the real-space Model~II with the size $20 \times 10 \times 10$ sites with a double bent domain wall in the $xy$ plane between domains with positive and negative signs of the bianisotropy parameter $\Omega = \pm 7$. The calculated spectrum of the eigenvalues, shown in Fig.~\ref{fig:Localization_zigzag}(a), demonstrates a bulk band gap in the range $-1.79 < \lambda < 4.56$. The bulk bands demonstrate delocalized eigenstates with different numbers of nodes and antinodes along the $x$-, $y$-, and $z$-directions, Fig.~\ref{sec:Zigzag}(b),(c). The eigenstates in the bandgap feature are localized along the double bent domain wall, Fig.~\ref{sec:Zigzag}(d)-(f). Similarly to the system with a planar interface considered in the main text, there are states with energies in the bulk bands with increased localization at the interface or boundary surfaces, Fig.~\ref{sec:Zigzag}(g),(h). Finally, interface-localized states are also observed in the bulk bands, with the example of such a mode shown in Fig.~\ref{sec:Zigzag}(i). It is seen that the obtained results for the two-domain system with a double-bent domain wall are similar to those calculated for the two-domain system with a planar interface.

\section{Numerical simulations of individual resonators}
\label{sec:Resonators}

\begin{figure*}[tbp]
    \centering
    \includegraphics[width=17cm]{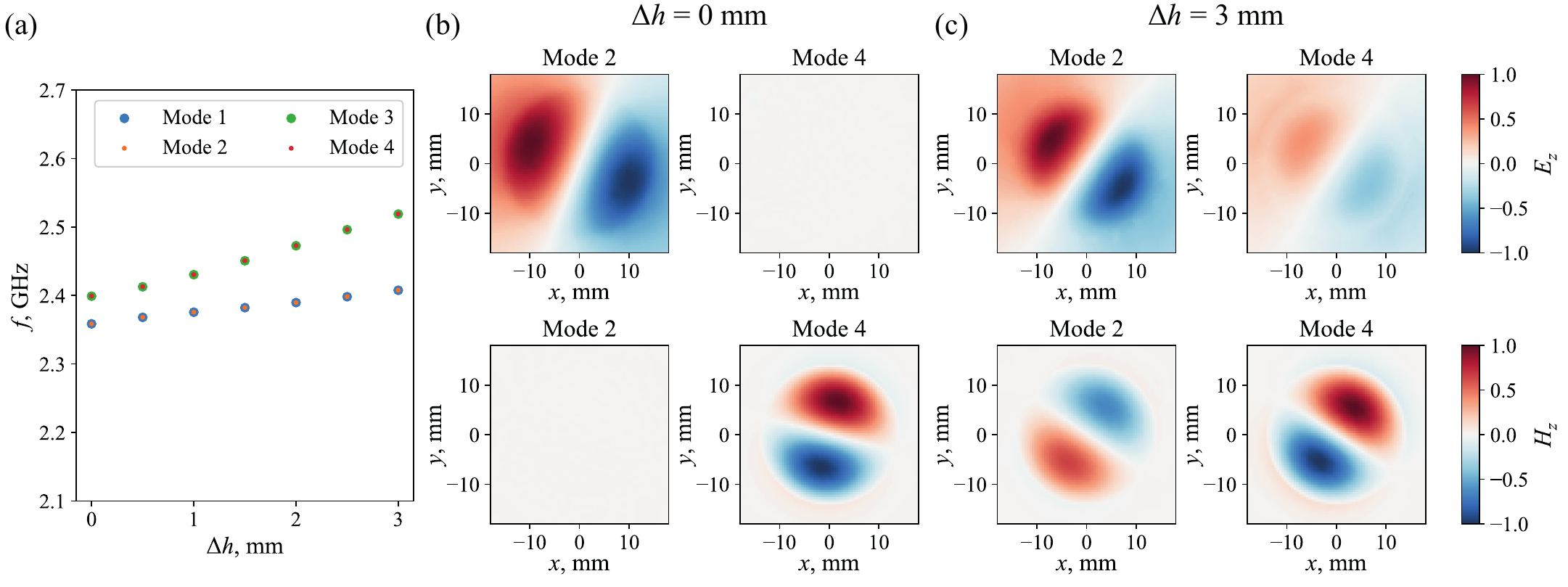}
    \caption{(a) Frequencies of the modes of the bianisotropic resonator formed by two cylinders with the heights $h_1=\Delta h$ and $h_2 = h-\Delta h$ for $h=12$~mm and different values of $\Delta h$. (b) Distribution of electric $E_z$ (top row) and magnetic $H_z$ (bottom row) fields in the $xy$ plane at $z = h/2$ corresponding to the second or fourth eigenmodes of a single resonator with $\Delta h=0$~mm. (c) The same as (b), but for the resonator with $\Delta h=3$~mm.}
    \label{fig:Resonator_Eigenmodes}
\end{figure*}

To study how the bianisotropic response of the resonator depends on its geometrical parameters, we perform numerical simulations of the eigenmodes of individual resonators with different ratios of the heights $h_{1}$ and $h_{2}$ of the cylinders that form the resonator. We change the height of the upper cylinder $h_{1}=\Delta h$ and of the lower cylinder $h_{2}=h-\Delta h$ in such a way that the height of the entire resonator $h=12$~mm remains constant.

Frequencies of the eigenmodes of the resonator for different values of $\Delta h$ are shown in Fig.~\ref{fig:Resonator_Eigenmodes}(a). For the considered $\Delta h$ in the range from $0$~mm to $3$~mm, the resonator supports a pair of doubly-degenerate eigenmodes. The frequencies of these modes, as well as their difference, increase with $\Delta h$. In the case of the cylindrical resonator with $\Delta h = 0$~mm, the lower pair of degenerate modes feature toroidal patterns of the vertical electric field component $E_{z}$ along with vanishing magnetic field $H_{z}$, that correspond to the magnetic dipole oriented in the $xy$ plane, as shown in the top panels of Fig.~\ref{fig:Resonator_Eigenmodes}(b). In contrast, the second pair of eigenmodes possess toroidal distributions of the magnetic field $H_{z}$, while the electric fields $E_{z}$ zeroes out, which is similar to the electromagnetic pattern created by the electric dipole located in the $xy$ plane, as shown in the bottom panels of Fig.~\ref{fig:Resonator_Eigenmodes}(b).

\begin{figure*}[t]
    \centering
    \includegraphics[width=8cm]{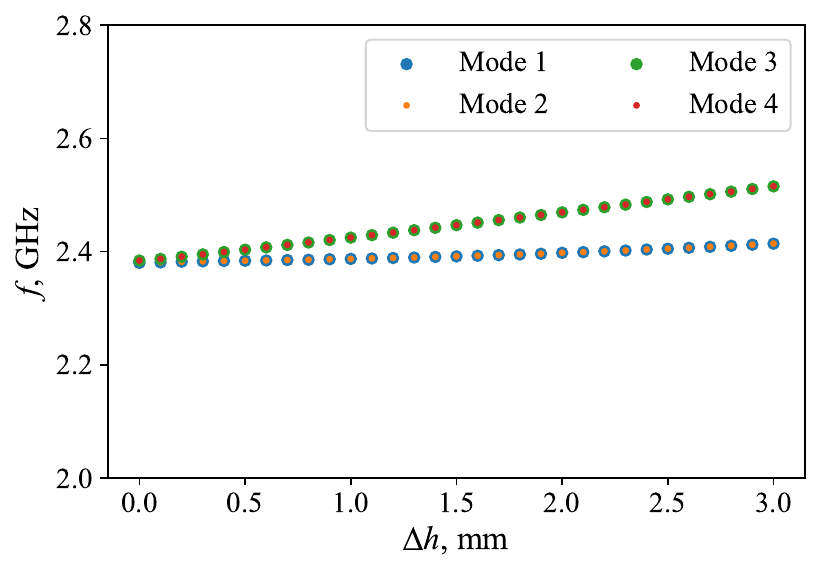}
    \caption{Frequencies of the bulk bands in the infinite lattice with spacings $a_{x,y}=45.1$~mm and $a_{z}=38$~mm at $M(\pi,\pi,0)$ high-symmetry point for different values of the height difference $\Delta h$. The heights of the upper and lower cylinders that form bianisotropic resonators are $h_1=\Delta h$ and $h_2 = h-\Delta h$ with $h=12$~mm, respectively.}
    \label{fig:Resonator_Bandgap}
\end{figure*}

The modes of the resonator with a broken inversion symmetry in the case of $\Delta h = 3$~mm simultaneously demonstrate features of the fields of electric and magnetic dipoles in the $xy$ plane for both pairs of doubly-degenerate modes, as shown in Fig.~\ref{fig:Resonator_Eigenmodes}(c). Such a hybridization highlights the electromagnetic coupling, i.e., a bianisotropic response. The electric and magnetic field profiles shown in Fig.~\ref{fig:Resonator_Eigenmodes}(b),(c) are extracted from the $xy$ plane at the midpoint of the resonator height $z=h/2$. Vertical components of the electric (magnetic) fields are normalized to the maximum value among all three components $E_x$, $E_y$, and $E_z$ ($H_x$, $H_y$, and $H_z$) in the considered plane.

Next, we perform numerical simulations of dispersion at $M(\pi, \pi, 0)$ high-symmetry point for the tetragonal lattice of resonators with different values of $\Delta h$ and periods $a_{x,y} = 45.1$~mm and $a_z = 38$~mm. The obtained results, shown in Fig.~\ref{fig:Resonator_Bandgap}, demonstrate a fourfold degeneracy for the lattice of cylindrical resonators with $\Delta h=0$~mm. However, this degeneracy is lifted, and the bulk band gap opens between the first and second pairs of doubly-degenerate modes as soon as $\Delta h$ is nonzero. The bandgap width increases with $\Delta h$. The results are in full agreement with those obtained for the considered Hamiltonians, where the bandgap width is proportional to the value of the bianisotropy parameter $\Omega$.

\section{Dynamic structure factor for different bianisotropy}
\label{sec:DSF_Omega}

To study the effects of the finite size of the considered system and the magnitude of the bianisotropy parameter $\Omega$ on the dynamic structure factor of the interface states, we consider Model~II as in the main text, but with different parameters.

\begin{figure*}[tbp]
    \centering
    \includegraphics[width=17cm]{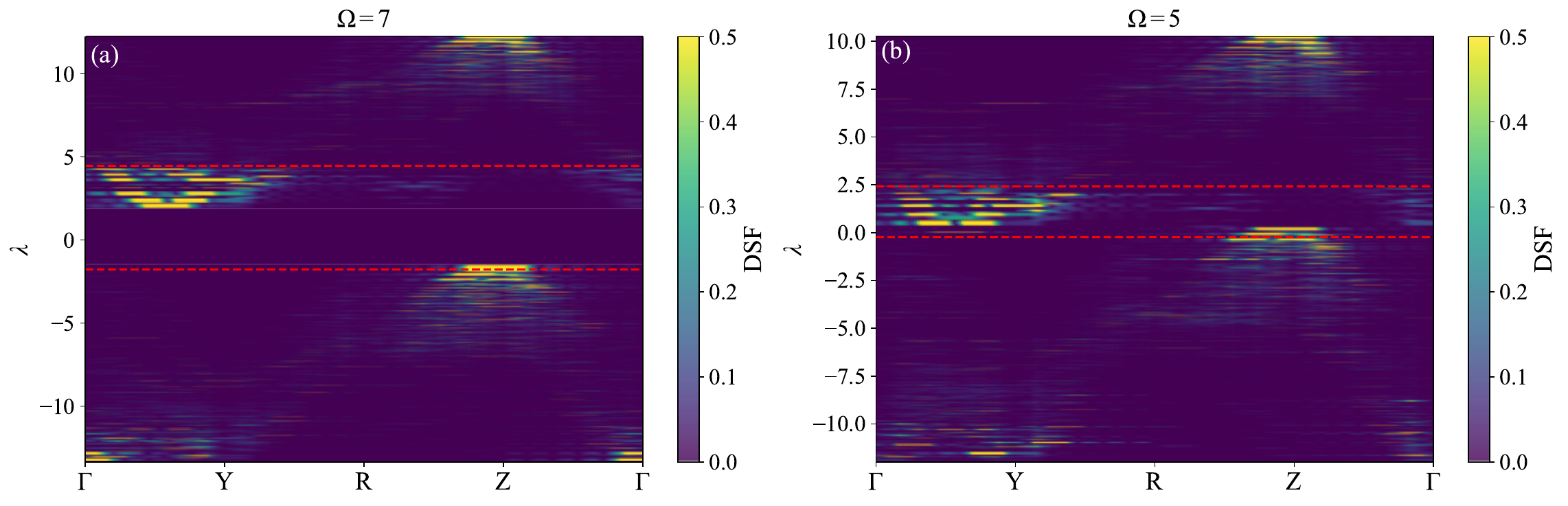}
    \caption{(a) Dynamic structure factor $S(\lambda, k_{y}, k_{z}) $ for the Model~II with bianisotropy parameter $\Omega = 7$ and the numbers of sites $20 \times 10 \times 10$ plotted along the trajectory $\Gamma$-$Y$-$R$-$Z$-$\Gamma$ in the Brillouin zone. Dashed horizontal lines denote the edges of the bulk band gap obtained as frequencies of the nearest bulk modes. Panel (b) is the same as (a), but for the system with bianisotropy parameter $\Omega = 5$.}
    \label{fig:DSF_Comparison}
\end{figure*}

Figure~\ref{fig:DSF_Comparison}(a) shows DSF $S(\lambda,k_{y},k_{z})$ for the Model~II composed of two domains with the sizes $10 \times 10 \times 10$ sites and bianisotropy parameters $\Omega=\pm 7$, respectively. It is seen that the broadening of the DSF distribution becomes more pronounced, highlighting its relation to the finite size of the system. The spacings between the energies of the interface states also increase due to the lower number of the modes in the system. However, the band gap edges do not demonstrate a significant change: for the model with the size $20 \times 20 \times 20$ sites their values are $\lambda_{\rm min}=-1.65$ and $\lambda_{\rm max}=4.43$, while for the model with $20 \times 10 \times 10$ sites we obtain $\lambda_{\rm min}=-1.77$ and $\lambda_{\rm max}=4.47$. At the same time, the energy of the lowest interface state demonstrates a larger change: from $\lambda=1.91$ for the size $20 \times 20 \times 20$ to $\lambda=2.06$ for the size $20 \times 10 \times 10$. Thus, it is possible that for a system of a sufficient size the energy of the bottom of the interface states band will reach the lower bulk band.

However, the idea that this gapped dispersion of the interface states results from the finite-size effects is contrasted by the results for the Model~II having the size $20 \times 10 \times 10$ sites with the values of bianisotropy parameters for two domains $\Omega=\pm 5$ shown in Fig.~\ref{fig:DSF_Comparison}(b). It is seen that for such a value of the bianisotropy parameter, the width of the bulk band gap becomes considerably lower, while the width of the interface states band changes insignificantly, and the interface states become gapless. The edges of the bulk band gap in this case are $\lambda_{\rm min}=-0.24$ and $\lambda_{\rm max}=2.4$, and the energy of the lowest in-gap interface state is $\lambda=0.37$. The dependence of the band gap crossing on the value of $\Omega$ is assumed for the trivial nature of the observed interface states.

%